\documentclass[
	a4paper, 
	10pt, 
	twoside, 
]{LTJournalArticle}

\usepackage{amssymb}
\usepackage{amsthm}
\usepackage{amsmath}
\usepackage{bm}
\usepackage{cleveref}
\usepackage{rotating}
\usepackage{tabularx}  
\usepackage{booktabs}
\usepackage{multirow}
\usepackage{multicol}
\usepackage{threeparttable}
\usepackage{subcaption}
\usepackage{graphicx}
\usepackage{setspace}

\newcommand\numberthis{\addtocounter{equation}{1}\tag{\theequation}}
\graphicspath{{Figures/}}

\runninghead{}

\footertext{} 

\title{Local Assortativity in Weighted and Directed Complex Networks} 

\author{%
    Marc Sabek \textsuperscript{1}\thanks{Corresponding author: \href{mailto:sabek@uni-wuppertal.de}{sabek@uni-wuppertal.de}} \and Uta Pigorsch\textsuperscript{1} 
}

\date{\footnotesize\textsuperscript{\textbf{1}}Schumpeter School of Business and Economics, University of Wuppertal, Gaußstraße 20, 42119, Wuppertal, NRW, Germany}

\renewcommand{\maketitlehookd}{%
	\begin{abstract}
		\noindent Assortativity, i.e. the tendency of a vertex to bond with another based on their similarity, such as degree, is an important network characteristic that is well-known to be relevant for the network's robustness against attacks. Commonly it is analyzed on the global level, i.e. for the whole network. However, the local structure of assortativity is also of interest as it allows to assess which of the network's vertices and edges are the most endangering or the most protective ones. Hence, it is quite important to analyze the contribution of individual vertices and edges to the network's global assortativity. For unweighted networks \citet{Piraveenan.2008, Piraveenan.2010} and \citet{Zhang.2012} suggest two allegedly different approaches to measure local assortativity. In this paper we show their equivalence and propose generalized local assortativity measures that are also applicable to weighted (un)directed networks. They allow to analyze the assortative behavior of edges and vertices as well as of entire network components. We illustrate the usefulness of our measures based on theoretical and real-world weighted networks and propose new local assortativity profiles, which provide, inter alia, information about the pattern of local assortativity with respect to edge weight. 
	\end{abstract}
\small\textit{\textbf{Keywords:} Network, local assortativity, third order graph metric, network robustness, assortativity profile} 
}

\begin{document}

\maketitle 


\section{Introduction}\label{chap_local_assortativity}

An important feature of a network's structure is its assortativity, which describes the tendency of vertices in the network to bond with other vertices that are similar with respect to a particular characteristic, predominantly vertex degree. Commonly assortativity is measured by the assortativity coefficient, which is defined as the Pearson correlation coefficient between the excess degrees of both ends of an edge. The assortativity coefficient has been originally proposed for undirected and unweighted networks by \cite{Newman.2001}, but extended to directed, see \cite{Newman.2003, Piraveenan.2012}, or weighted networks, see \cite{Leung.2007}. \cite{U.Pigorsch.2022} propose a generalized assortativity coefficient that nests the other ones, but allows to measure assortativity also for directed and weighted networks where weights are incorporated using excess vertex strength rather than excess vertex degree. 

All of these assortativity coefficients are global, in the sense that they measure the overall assortativity of the network. Hence, they do not provide any information about the assortative mixing behavior of individual vertices, edges, groups or communities within the network. However, this information is of particular importance for the robustness of the overall network. In fact, an assortative (and also disassortative) network may exhibit both, assortative as well as disassortative components, such that shocks to individual components may have a different impact on the overall resilience of the network. Thus, knowing the local structure of assortativity can help to understand which components, like vertices or edges are particularly stability-threatening and which are stability-protective. This, in turn may be useful for developing efficient strategies for both breaking up a network (e.g., with vaccination in disease spreading social networks) and for protecting particularly vulnerable networks (e.g., financial networks or technological networks, such as the Internet), cf. \cite{Newman.2002}.

Moreover, in contrast to global assortativity, local assortativity can be considered as a \textit{third-order} graph metric providing additional information, e.g. allowing for further differentiation in graphs that have the same degree distribution (\textit{first-order} metric) as well as the same assortativity structure (\textit{second-order} metric), cf. \cite{Noldus.2015}. As an example consider the \textit{Erd\H{o}s-Rényi random graph} (ER) and Barabási-Albert (BA) models. Both are well-known to generate globally non-assortative networks, see \cite{Newman.2002}, however, only the ER model generates graphs that also show local non-assortative tendencies, while in the BA model vertices that join the network at an earlier point in time tend to be locally disassortative.

The importance of local assortativity has already been recognized in \cite{Piraveenan.2008} and \cite{Zhang.2012}, who propose measures of local assortativity in unweighted networks with the latter being an edge-based measure and the former a vertex-based measure. In this paper we first show that these two measures, that are usually treated as being two distinct concepts of local assortativity, are equivalent up to a scaling factor. Secondly, as these measures are defined only for unweighted networks, but many real-world networks exhibit weighted edges, we propose generalized local assortativity measures that are applicable to unweighted networks, but importantly, also to weighted networks, which can be either undirected or directed. Moreover, our new measures nest the previously proposed local assortativity measures as special cases and are flexible in the sense that they can be either vertex- or edge-based. 

The remainder of this paper is structured as follows: \Cref{sec_background} provides a review of the related literature on local assortativity in unweighted and undirected networks and demonstrates the equivalence of the existing two allegedly different measures of local assortativity. Building on the corresponding unifying framework, we introduce our generalized local assortativity measures in \Cref{sec_generalised_local_assortativity}. The application and usefulness of these generalized local assortativity measures are illustrated based on simulated networks and two real-world networks. In our empirical analysis, we also compute local assortativity profiles, which are very helpful for characterizing the local assortativity pattern of a network. \Cref{sec_discussion_local} discusses the results and gives suggestions for future research.

\section{Background and Related Literature}\label{sec_background}
We begin our review on the related literature by first defining the global assortativity coefficient of \cite{Newman.2002}. To this end, let $p(k^\prime)=p_{k^\prime}$ denote the degree distribution, i.e., the probability of a randomly chosen vertex having degree $k^\prime$; $q(k)=q_{k}$ denotes the excess (or remaining) degree distribution, i.e., the probability that a vertex reached by following a randomly chosen edge is connected to $k$ other vertices (excluding the vertex associated with the randomly chosen edge). Since $k = k^\prime - 1$, excess degree is distributed proportional to $(k+1)p_{k+1}$ and the normalized excess degree distribution $q_{k}$ is given by:
\begin{align*}
	q_{k} = \frac{(k+1)p_{k+1}}{\sum_{k^\prime} k^\prime p_{k^\prime}},
\end{align*}
where $ \sum_{k^\prime} k^\prime p_{k^\prime} $ is the expected degree of a vertex. 

Furthermore, let $e_{j k}$ denote the joint distribution of excess degrees of either end of an edge, i.e., the probability that a randomly chosen edge connects two vertices with excess degrees $ j $ and $k$, see e.g. \cite{Callaway.2001}. Then, the global assortativity coefficient $r$ proposed in \cite{Newman.2002} is defined by
\begin{align}
	r & = \frac{1}{\sigma^2_{q_k}}\sum_{jk}jk(e_{jk} - q_jq_k) \numberthis\label{eq_newman_assortativity}   
\end{align}
with $\sigma^2_{q_k}$ denoting the variance of the distribution $q_k$. $r$ is the Pearson correlation coefficient between the excess degrees at either end of an edge. Networks with $0<r\le 1$ are called assortative, those with $-1\le r<0$ disassortative, and those with $r=0$ non-assortative. The coefficient can also be expressed as 
\begin{align*}
	r & = \frac{1}{\sigma^2_{q_k}}\Big[\sum_{jk}jke_{jk} - U_{q_k}^2\Big]
\end{align*}
where $U_{q_k}=\sum_k k q_k$ denotes the expected value of the excess degree of an end of an edge.

Based on this assortativity coefficient \cite{Piraveenan.2008, Piraveenan.2010} derive a \textit{vertex-based} local measure of assortativity that measures the contribution of an individual vertex to the overall (global) assortitivity of the network $r$. The resulting local assortativeness $\rho_v$ of a vertex $v$ of a network with $M$ edges is given by
\begin{align}
	\rho_v=\frac{k^\prime_v (j_v\bar{k}_u - {U}_{q_k}^2)}{2M{\sigma}^2_{q_k}},\numberthis\label{eq_piraveenan_theoretical}
\end{align}
where, $k^\prime_v$ is the total degree of vertex $v$, $j_v$ its excess degree, and $\bar{k}_u$ the average excess degree of its neighbors $u\neq v$. The scaling by $2M$ ensures that the sum of the local assortativeness of all $N$ vertices of the network equals the global assortativity coefficients, i.e. that
\begin{align*}
	r=\sum_{i=1}^{N}\rho_i.
\end{align*}
For practical purposes local assortativeness of a vertex is computed by replacing ${U}_{q_k}$ by the corresponding sample mean $\bar{U}_{q_k}=\frac{1}{M}\sum_{e=1}^M \frac{1}{2} (j_e + k_e)$,  where $j_e$ and $k_e$ are the excess degrees of the ends of edge $e$, and ${\sigma}^2_{q_k}$ by the corresponding sample variance $\hat{\sigma}^2_{q_k}=\sqrt{\frac{1}{M}\sum_{e=1}^M \frac{1}{2} (j_e^2 + k_e^2) - \bar{U}_{q_k}^2}$.\footnote{In an undirected network, each edge $e$ has two ends $j_e$ and $k_e$, and, thus, there are $2M$ ends in total. Therefore, the mean excess degree of an end of an edge is obtained by averaging the mean of the excess degrees of the ends of an edge over all edges of the network. Similarly for the computation of the standard deviation of the excess degree of an end of an edge.} We refer to this as the \textit{vertex assortativeness value} of a vertex $v$:
\begin{align}
	\rho_v=\frac{k^\prime_v (j_v\bar{k}_u - \bar{U}_{q_k}^2)}{2M\hat{\sigma}^2_{q_k}}.\numberthis\label{eq_piraveenan}
\end{align}
A vertex is called assortative if its contribution to the network's global assortativity is positive, $\rho_v>0$, disassortative if its contribution is negative, $\rho_v<0$, and non-assortative, if it has no contribution, $\rho_v=0$.

As an alternative, allegedly different approach \cite{Zhang.2012} introduced an \textit{edge-based} measure of local assortativity. To this end, the authors first rewrite the global assortativity coefficient given in \cref{eq_newman_assortativity} as\begin{align}
r& = \frac{E\big[(J-U_{q_k})(K-U_{q_k})\big]}{\sigma^2_{q_k}},\numberthis \label{eq_local_assortativity_uac}
\end{align}
where $J$ and $K$ are the variable excess degrees of the ends of an edge. Based on this representation they propose to measure the contribution of each individual edge $e$ to the overall assortativity of the network $r$ by
\begin{align}
\rho_e = \frac{(j_e-{U}_{q_k})(k_e-{U}_{q_k})}{M\sigma^2_{q_k}}.\label{eq_local_assortativity_uac_final_theoretical}
\end{align}
In practice, the theoretical quantities in \cref{eq_local_assortativity_uac_final_theoretical} need to be estimated by their sample counterparts, resulting in, what we refer to as the \textit{edge assortativeness value} of edge $e$:
\begin{align}
\rho_e = \frac{(j_e-\bar{U}_{q_k})(k_e-\bar{U}_{q_k})}{M\hat{\sigma}^2_{q_k}}.\label{eq_local_assortativity_uac_final}
\end{align}
These assortativeness values can be interpreted as an \textit{edge-based} local assortativity measure. In particular, an edge is considered assortative if its contribution to the global assortativity coefficient is positive, i.e. $\rho_e > 0$, and disassortative, in case its contribution is negative, i.e., $\rho_e < 0$ and, obviously, non-assortative, if it does not contribute.

Defining local assortativity on an edge basis is advantageous, since this allows to determine the assortativity of any set of edges, $E_{\text{target}}$, that may be of interest, by computing the sum of the edge assortativeness values of all the edges in that set, i.e.\begin{align*}\rho^{\text{UAC}} = \sum_{e \in E_{\text{target}}} \rho_e. \numberthis \label{eq_uac}
\end{align*}
\cite{Zhang.2012} refer to this as the \textit{universal assortativity coefficient} (UAC).
Setting the target edge set $E_{\text{target}}$ in \cref{eq_uac} to the entire edge set $E$ consisting of the $M$ edges of the network yields the assortativity coefficient of \cite{Newman.2002}, i.e. $\sum_{e=1}^M\rho_e = r$.\footnote{Note, that, in undirected networks, the global assortativity coefficient in \cref{eq_local_assortativity_uac} can be estimated by:
\begin{align*}
	r = \frac{\frac{1}{M}\sum_{e=1}^M (j_e - \bar{U}_{q_k})(k_e - \bar{U}_{q_k})}{\hat{\sigma}^2_{q_k}}.\numberthis\label{eq_assortativity_estimated}
\end{align*}
}

Moreover, \cite{Zhang.2012} argue that the assortativeness values $\rho_e$ can be used to easily derive an \textit{edge-based} local assortativity measure $\rho_v$ for some vertex $v$ by setting the target edge set $E_{\text{target}}$ to the edges emanating from that vertex $v$. More precisely, for undirected networks this gives:
\begin{align}
	\rho_v = \sum_{u=1}^n \rho_{e_{vu}},
	\label{eq_vertex_local_assortativity}
\end{align}
where $\rho_{e_{vu}}$ is the edge assortativeness value of the edge $e$ that has end vertices $v$ and $u$. A vertex is considered again as assortative if $\rho_v > 0$, disassortative if $\rho_v < 0$ and non-assortative, if $\rho_v=0$.

The summation of $\rho_v$ for all vertices yields:\footnote{Note that the computation of $\rho_v$ in \cref{eq_vertex_local_assortativity} is similar to the computation of the vertex degree $k^\prime_v$ of vertex $v$ in an undirected network, however, instead of summing the elements of the network's adjacency matrix $\bm{A} = [a_{vu}]$, the sum is computed over the local edge assortativeness values of connected vertices. Thus, the sum over the edge assortativeness values of all vertices in the network is $2r$ as compared to the sum of all vertex degrees, which is $2M$.}
\begin{align*}
	\sum_{v=1}^n \rho_v = \sum_{uv} \rho_{e_{uv}} = 2r.\numberthis\label{eq_rho_v}
\end{align*}

Obviously, both approaches to local assortativity, that of \cite{Piraveenan.2008} and that of \cite{Zhang.2012}, allow to derive a measure of the assortativity of a vertex $v$. In the following, we show that both measures, although based on different concepts, are equivalent. To this end reconsider \cref{eq_vertex_local_assortativity}. Replacing the right-hand side of the equation by its definition in \cref{eq_local_assortativity_uac_final} we obtain:
\begin{align*}
	\rho_v = \sum_{u=1}^n \rho_{e_{vu}}
	 & = \sum_{u=1}^n \frac{(j_v-\bar{U}_{q_k})(k_u-\bar{U}_{q_k})}{M\hat{\sigma}^2_{q_k}}                                \\
	 & = \sum_{u=1}^n \frac{(j_vk_u - \bar{U}_{q_k}^2)}{M\hat{\sigma}^2_{q_k}} \numberthis\label{eq_first}                \\
	 & =  \frac{\sum_{u=1}^n j_v k_u - \sum_{u=1}^n \bar{U}_{q_k}^2}{M\hat{\sigma}^2_{q_k}}                               \\
	 & = \frac{ j_v \sum_{u=1}^n k_u - \sum_{u=1}^n \bar{U}_{q_k}^2}{M\hat{\sigma}^2_{q_k}}                               \\
	 & = \frac{ j_v k^\prime_v \bar{k}_u - k^\prime_v \bar{U}_{q_k}^2}{M\hat{\sigma}^2_{q_k}}\numberthis\label{eq_second} \\
	 & = \frac{  k^\prime_v (j_v\bar{k}_u - \bar{U}_{q_k}^2)}{M\hat{\sigma}^2_{q_k}},\numberthis\label{eq_third}
\end{align*}
where \cref{eq_first} is an equivalent transformation, cf. \cite[eq. (4)]{Newman.2002}.
\Cref{eq_second} holds as the number of neighbours of a vertex corresponds to its degree, and, thus, $\frac{1}{k^\prime_v}\sum_{u=1}^n k_u = \bar{k}_u \Leftrightarrow \sum_{u=1}^n k_u = k^\prime_v \bar{k}_u$, where $\bar{k}_u$ is again the mean excess degree of the neighbours of vertex $v$.
Finally, \cref{eq_third} shows that the vertex-based local assortativity measure of \cite{Zhang.2012}, $\rho_v$, is twice the local assortativity measure proposed by \cite{Piraveenan.2008}, see \cref{eq_piraveenan}.

This demonstrates that both definitions of local assortativity, the one by \cite{Zhang.2012} and the one by \cite{Piraveenan.2008} are equivalent up to a constant scaling factor. Note that \cite{Piraveenan.2008} scale their local measure, such that summing over the local vertex assortativeness values yields the assortativity coefficient $r$. This, however, is a matter of choice, and is achieved by multiplying the denominator in \cref{eq_third} by $2$, because $\sum_{v=1}^n \rho_v = 2r \Leftrightarrow r = \sum_{v=1}^n \frac{\rho_v}{2}$. 

\cite{Piraveenan.2010} propose a slightly modified definition of local assortativity, where the numerator of \cref{eq_third} is given by $j_v k_v^\prime(\bar{k}_u - \bar{U}_{qk})$. However, in order to derive this definition \cite{Piraveenan.2010} had to make a questionable assumption on how to split the contribution to the global assortativity among vertices, see \cite[eq. (15) in Appendix A]{Piraveenan.2010a}. More precisely, when determining the contribution of a vertex $i$ to the term $\mu_q$ (which corresponds to $\bar{U}_{q_k}$ using our notation), the arising cross-terms, $2k_ik_j$, are divided equally among vertices $i$ and $j$, disregarding the potentially different excess degrees, $k_i$ and $k_j$, respectively. The definition of \cite{Piraveenan.2008}, does not rely on this additional assumption. Moreover, the resulting local vertex assortativeness of \cite{Piraveenan.2010} would have to be additionally scaled, to ensure that they sum to the global assortativity coefficient. This can be seen from a comparison to the definition in \cref{eq_third}, which satisfies this equality. Unless $\bar{U}_{qk} = 0 \vee j_v \neq 0$ or $\bar{U}_{qk} = j_v \neq 0$ both definitions are different, in general. Therefore, we consider the author's first measure of vertex assortativity, see \cite{Piraveenan.2008}, to be more appropriate. Also because it can be derived based on the alternative edge-based approach to local assortativity of \cite{Zhang.2012}.

In the remainder of the paper we, thus, build on the approach of \cite{Zhang.2012}. It is especially appealing, as it allows to assess the contribution of both, edges and vertices (and groups thereof) to the global assortativity coefficient and, as such, provides a more detailed description of the local assortativity structure of the network. Moreover, based on the assortativeness values, $\rho_e$ and $\rho_v$, the following additional measures can be derived that allow to further differentiate the local connectivity tendencies of a network: The relative frequency of assortative edges, $P(\rho_e>0)$ and similarly, the relative frequency of assortative vertices, $P(\rho_v >0)$. \cite{Zhang.2012} also consider $\overline{(\rho_e)_{+}}$, and $\overline{(\rho_e)_{-}}$ which are defined as:
\begin{align*}
    \overline{(\rho_e)_{+}} &= \frac{1}{\#\{\rho_e>0\}} \sum_{\rho_e > 0} \rho_e,
\end{align*}
and
\begin{align*}
    \overline{(\rho_e)_{-}} &= \frac{1}{\#\{\rho_e<0\}} \sum_{\rho_e < 0} \lvert\rho_e\rvert.
\end{align*}
The authors refer to these values as the \textit{average assortative (or disassortative) strength} of an edge. However, as this terminology might be misleading in the context of weighted networks, we instead refer to these measures as \textit{mean absolute magnitude of assortative edges} and \textit{mean absolute magnitude of disassortative vertices}, respectively.

As pointed out in \cite{Zhang.2012}, all these measures can also be applied to directed networks, but further details are not provided by the authors. In \cite{Piraveenan.2012} the vertex-based measure of local assortativity is explicitly extended to directed networks. However, to the best of our knowledge, the measurement of local assortativity in weighted networks has not been considered in the literature yet. Therefore, in the following section, we propose a generalized approach to measuring local assortativity in weighted networks.

\section{Generalized Local Assortativity}\label{sec_generalised_local_assortativity}

In weighted complex networks, vertex degree generalizes to vertex strength. \cite{Barrat.2004} defines the strength of a vertex $u$ to be the total weight of its connections, i.e., $s^\prime_u = \sum_{v\in V} w_{uv}$ where $V$ is the vertex set and $w_{uv}$ is the weight of the edge that connects vertices $u$ and $v$.
Thus, it is sensible to also generalize the definition of the assortativity coefficient, $r$, in \cref{eq_newman_assortativity} by incorporating excess vertex strengths rather than excess vertex degrees in weighted networks. This allows for deriving a weighted local assortativity measure related to the global generalized assortativity coefficient. In the following, we first focus on weighted, but undirected networks and introduce the corresponding generalized local assortativity measures. Thereafter, we give a detailed description of its extension to directed networks.

\subsection{Undirected Weighted Networks}

Let $p_{s^\prime}$ denote the strength distribution, i.e., the probability that a randomly chosen vertex has \mbox{strength $s^\prime$} and
$q_{s}$ the excess strength distribution, i.e., the probability that a vertex reached by following a randomly chosen edge of weight $\omega$ has excess strength $s$.
Note that the excess strength $s$ of a vertex depends on the weight $\omega$ of the edge we arrived along, i.e., $s = s^\prime - \omega$, and, hence, is distributed proportional to $(s+\omega)p_{s+\omega}$ with normalized distribution given by:
\begin{align*}
	q_{s} = \frac{(s+\omega)p_{s+\omega}}{\sum_{s^\prime} s^\prime p_{s^\prime}},
\end{align*}
where $ \sum_{s^\prime} s^\prime p_{s^\prime} $ is the expected strength of a vertex. We denote by $e_{st}$ the joint distribution of excess strengths of either end of an edge, i.e., the probability that a randomly chosen edge connects two vertices with excess strengths $s$ and $t$. Similarly to \cref{eq_newman_assortativity}, we can, thus, define global assortativity based on weighted vertex values, i.e. excess strength:
\begin{align*}
	r^\omega & = \frac{1}{\sigma^2_{q_s}}\sum_{st}st(e_{st} - q_s q_t),\numberthis\label{eq_r_omega}
\end{align*}
which is equivalent to:
\begin{align*}
	r^\omega & = \frac{E\big[(S-U_{q_s})(T-U_{q_s})\big]}{\sigma^2_{q_s}},\numberthis\label{eq_r_prime}
\end{align*}
where $S$ and $T$ are the excess strengths of the ends of an edge and $U_{q_s}$ denotes the mean of the distribution $q_s$, which is the mean excess strength of an end of an edge, and $\sigma^2_{q_s}$ is the variance of the distribution $q_s$.

In order to quantify the amount of assortative mixing \cite{Newman.2002} follows \cite{Callaway.2001} by employing the (unweighted) connected degree-degree correlation function. This is suitable for \textit{unweighted} networks. However, in \textit{weighted} networks we must consider information on the weights of the observational pairs $s_e$ and $t_e$ which is provided by the weight $\omega_e$ of the connecting edge $e$. This information can be incorporated into the correlation coefficient between weighted vertex values (\cref{eq_r_prime}) by estimating the theoretical quantities by their \textit{weighted} sample counterparts (again by averaging over the edges of the network), see \cite{Price.1972}. The resulting global assortativity coefficient for weighted networks, also called weighted assortativity coefficient, is the weighted correlation coefficient between the weighted vertex values of the ends of an edge, as introduced in \citep{U.Pigorsch.2022}:
\begin{align*}
	r^\omega = \frac{\frac{1}{H}\sum_{e=1}^M \omega_e(s_e - \bar{U}^{\omega}_{q_s})(t_e - \bar{U}^{\omega}_{q_s})}{(\hat{\sigma}^{\omega}_{q_s})^2},\numberthis\label{eq_r_prime_estimated}
\end{align*}
where $s_e$ and $t_e$ are the excess strengths of the ends of edge $e$, $H = \sum_{e=1}^M \omega_e$ denotes the sum of edge weights; \mbox{$\bar{U}^{\omega}_{q_s} = H^{-1}\sum_{e=1}^M \frac{1}{2}\omega_e(s_e + t_e)$} the weighted sample mean, and \mbox{$\hat{\sigma}^{\omega}_{q_s} = \sqrt{H^{-1}\sum_{e=1}^M \frac{1}{2}\omega_e\big(s_e^2 + t^2_e\big) - (\bar{U}^{\omega}_{q_s})^2}$} the weighted sample standard deviation of the excess strength of an end of an edge.

Based on this weighted assortativity coefficient, we can derive the \textit{weighted edge assortativeness values}:
\begin{align*}
	\rho^\omega_e = \frac{\omega_e(s_e - \bar{U}^{\omega}_{q_s})(t_e - \bar{U}^{\omega}_{q_s})}{H(\hat{\sigma}^{\omega}_{q_s})^2}.\numberthis\label{eq_rho_weighted}
\end{align*}
Edges are considered to be disassortative or assortative if their contribution to the global weighted assortativity coefficient $r^\omega$ is negative, i.e. $\rho^\omega_e < 0$, or positive, i.e. $\rho^\omega_e > 0$, respectively, and non-assortative if $\rho^\omega_e = 0$.

We aim at a generalization of the local assortativity measure and, therefore, follow \citep{U.Pigorsch.2022} and additionally introduce the parameters $\alpha, \beta \in \{0,1\}$. The parameters allow to control the degree of generalization of the local assortativity measures. More precisely, the parameter $\alpha$ controls whether excess degrees ($\alpha=0$) or excess strengths ($\alpha=1$) are used as vertex values, and $\beta$ determines whether the unweighted ($\beta=0$) or weighted ($\beta=1$) correlation between the vertex values of the ends of an edge is computed.
Moreover, due to these tuning parameters previous (local) assortativity measures are nested as special cases, as we will detail in the following.
To this end, let $s^\ast_u = \sum_{v\in V} w_{uv}^\alpha$ be a modified version of vertex strength $s^\prime_u$.
Clearly, if $\alpha = 1$, then $s^\ast$ equals the vertex strength, i.e., $s^\ast = s^\prime$, whereas, if $\alpha = 0$, then $s^\ast$ reduces to ordinary vertex degree, i.e., $s^\ast = k^\prime$.
Furthermore, let $\Omega = \sum_{e=1}^M \omega_e^\beta$. Obviously, if $\beta = 1$, then $\Omega = H$, i.e., $\Omega$ corresponds to the sum of edge weights, whereas, if $\beta = 0$, then $\Omega = M$, i.e., $\Omega$ equals the number of edges in the network. Based on these parameters, $(\alpha, \beta)$, we define the corresponding \textit{generalized edge assortativeness values} as:
\begin{align*}
	\rho_e^{\omega}(\alpha,\beta) = \frac{\omega_e^\beta\big[l_e-\bar{U}^{\omega}_{q_s}(\alpha, \beta)\big]\big[m_e-\bar{U}^{\omega}_{q_s}(\alpha, \beta)\big]}{\Omega\big[\hat{\sigma}^{\omega}_{q_s}(\alpha, \beta)\big]^2},\numberthis\label{eq_rho_weighted_alpha_beta}
\end{align*}
where $l_e$ and $m_e$ are the vertex values of the ends of edge $e$,
e.g. $l_e = s^\ast_l - \omega_e^\alpha$; the (weighted) sample mean and standard deviation as functions of the parameters $(\alpha, \beta)$ are defined by $\bar{U}^{\omega}_{q_s}(\alpha, \beta) = \Omega^{-1}\sum_{e=1}^M \frac{1}{2}\omega_e^\beta(l_e + m_e)$ and
$\hat{\sigma}^{\omega}_{q_s}(\alpha, \beta) = \sqrt{\Omega^{-1}\sum_{e=1}^M \frac{1}{2}\omega_e^\beta \big(l_e^2 + m^2_e\big) - \big[\bar{U}^{\omega}_{q_s}(\alpha, \beta)\big]^2}$, respectively.
The expression in \cref{eq_rho_weighted_alpha_beta} is the most general version of the (weighted) assortativeness values (for undirected networks) and nests the previous (local) assortativity measures as follows:
for $\alpha = 0$ and $ \beta = 0$, the values $\rho^\omega_e(\alpha, \beta)$ reduce to the original edge assortativeness values of \cite{Zhang.2012} (see \cref{eq_local_assortativity_uac_final}), i.e. $\rho^\omega_e(0, 0) = \rho_e$, such that, $\sum_{e=1}^M \rho^\omega_e(0,0) = r$.
Whereas, for $\alpha = 1$ and $\beta = 1$, the values equal the weighted edge assortativeness values (see \cref{eq_rho_weighted}), i.e. $\rho^\omega_e(1,1) = \rho^\omega_e$, and, thus, $\sum_{e=1}^M \rho^\omega_e(1,1) = r^\omega$.
Generally, summing the values $\rho^\omega_e(\alpha, \beta)$ for all edges $M$ in the network yields the \textit{generalized assortativity coefficient} of \cite{U.Pigorsch.2022}, $r^\omega(\alpha, \beta)$, i.e., $\sum_{e=1}^M \rho^\omega_e(\alpha, \beta) = r^\omega(\alpha, \beta)$.

Based on the generalized edge assortativeness values we can define \textit{generalized vertex assortativeness values} by summing generalized edge assortativeness values over the edges emanating from a vertex $v$:
\begin{align*}
\rho^\omega_v(\alpha, \beta) = \sum_{u = 1}^n \rho^\omega_{e_{vu}}(\alpha, \beta). 
\end{align*}
Considering an arbitrary target edge set $E_{\text{target}}$ when summing over the edge assortativeness values yields a generalized version of the UAC of \cite{Zhang.2012}, which we refer to as the \textit{generalized universal assortativity coefficient} (GUAC):
\begin{align}
	\rho^{\text{GUAC}}(\alpha, \beta)=\sum_{e\in E_{\text{target}}} \rho_e^\omega(\alpha, \beta).\label{eq_guac}
\end{align}
The GUAC is a versatile coefficient that, e.g., can be used to determine the contribution of either a set of vertices or edges to the global assortativity. It is up to the researcher to choose which vertices and edges are interesting to consider. Possible interesting vertices can include the top $n$ most assortative or most disassortative vertices (or edges). However, it can also be interesting to examine the vertices that can be combined into a community (or the edges connecting them) with regard to their aggregated assortativity. Another interesting example, which is considered in the following, is the determination of the assortativity of isomorphic components in a network.

\begin{figure}[!htb]
	\centering
	\includegraphics[width=\linewidth]{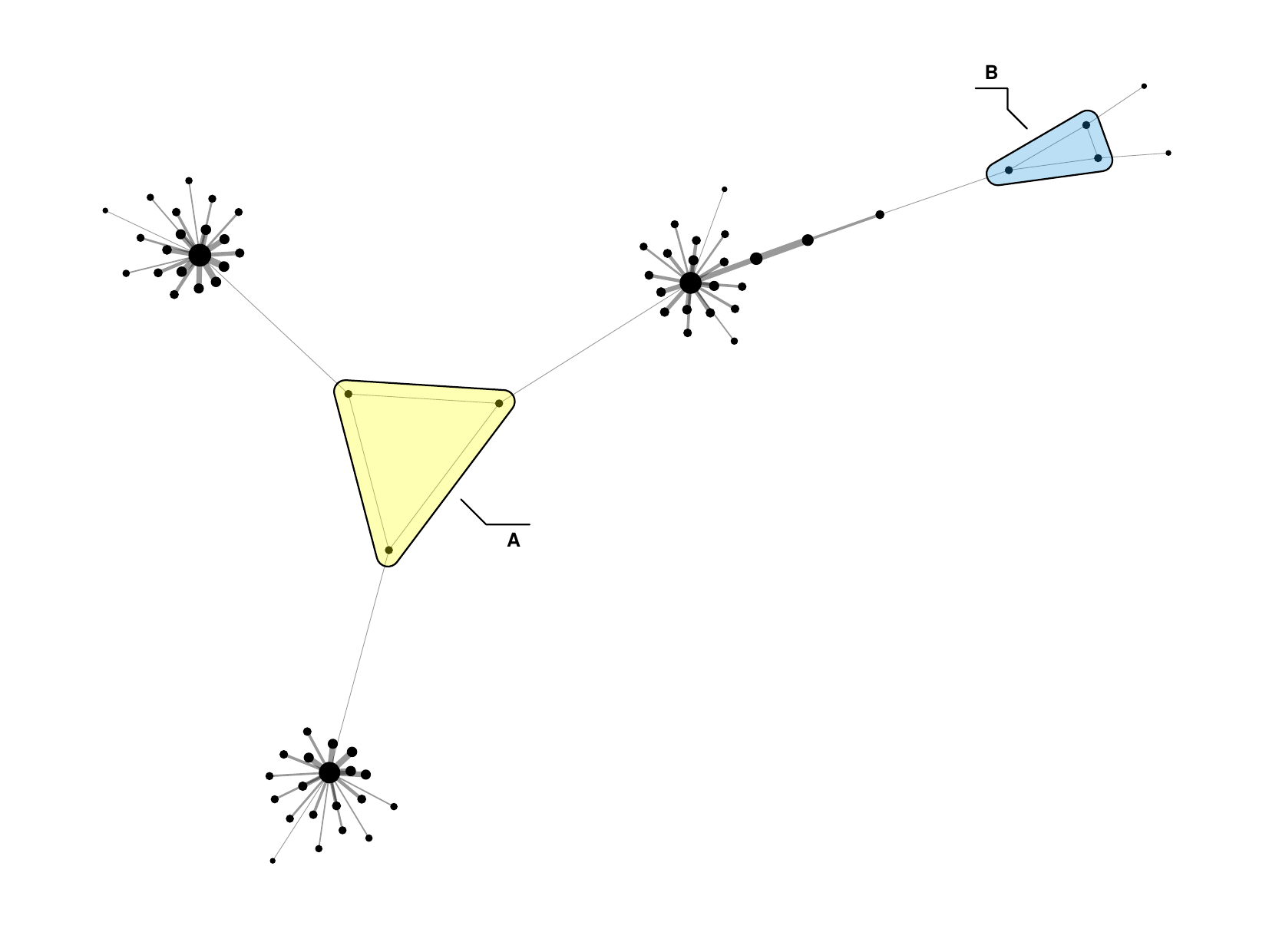}
	\caption[Differences of local connectivity patterns of isomorphic components.]{\textbf{Differences of local connectivity patterns of isomorphic components (A and B).} The depicted weighted undirected graph with $n = 70$ vertices and $M= 71$ edges is based on the one presented in \cite{Zhang.2012}, but is complemented by randomly assigned edge weights (except for the edges that are incident with vertices of components $A$ or $B$, which have the same weight of one). Higher edge weights are indicated by thicker edges. The depicted size of vertices is proportional to the vertex' total vertex strength.}
	\label{fig_homomorphic_components_local_assortativity}
\end{figure}

\Cref{fig_homomorphic_components_local_assortativity} shows the capability of the GUAC to identify the differences of weighted local connectivity patterns of isomorphic components. The example weighted undirected graph with $n = 70$ vertices and $M= 71$ edges is based on the one presented in \cite{Zhang.2012}, but is complemented by randomly assigned edge weights (drawn from a uniform distribution over the interval [1,10], except for the edges that are incident with vertices of components $A$ or $B$, which have weight one). The network is overall disassortative with respect to degrees, as $r^\omega(0,0) = -0.800$, and also (even slightly more) overall disassortative with respect to strengths, as $r^\omega(1,1) = -0.828$. However, the components $A$ and $B$ show different local connectivity patterns in the sense of how they connect to the rest of the network. In order to quantify this behaviour the GUAC of the components is computed by summing the local edge assortativeness values that connect each component to the rest of the network, respectively. Although they are isomorphic components, $A$ tends to connect degree disassortatively to the rest of the network, $\rho^{\text{GUAC}}_{A}(0,0) = -0.034$, whereas $B$ forms degree assortative connections, $\rho^{\text{GUAC}}_{B}(0,0) = 0.025$. This can be verified in \Cref{fig_homomorphic_components_local_assortativity}, where the component $A$, of which the vertices have a relatively low degree, is surrounded by high-degree vertices, whereas component $B$ is connected to also low degree vertices. If, however, edge weights are considered, then, the components $A$ and $B$ are almost non-assortative, as $\rho^{\text{GUAC}}_{A}(1,1) = -0.009$ and $\rho^{\text{GUAC}}_{B}(1,1) = 0.007$. The signs of $\rho^{\text{GUAC}}_{A}(1,1)$ and $\rho^{\text{GUAC}}_{B}(1,1)$ still differ, because of the way each component connects to the rest of the network, i.e., $A$ is surrounded by vertices with high strengths, whereas $B$ connects to low strength vertices. However, the contribution to the overall disassortativity of the network is equally small for both components $A$ and $B$. This is because of the comparatively low weights of the edges that connect the components to the rest of the network (all equal one).

As a result, the components $A$ and $B$ show different local connectivity patterns with respect to their degrees, but similar local connectivity patterns with respect to their strengths. Thus, GUAC allows a further distinction between the local connectivity patterns of the components and their contribution to the global assortativity of the network when edge weights are additionally considered.

Moreover, with the help of both generalized edge and vertex assortativeness values, generalizations of the unweighted measures can be derived, that allow for further differentiation of the local connectivity tendencies of a weighted network, such as the proportion of assortative edges, $P\big(\rho^\omega_e(\alpha, \beta) > 0\big)$, the proportion of assortative vertices, $P\big(\rho^\omega_v(\alpha, \beta) > 0\big)$, the mean absolute magnitude of assortative edges $\overline{\big(\rho^\omega_e(\alpha,\beta)\big)_+}$ and the mean absolute magnitude of disassortative edges, $\overline{\big(\rho^\omega_e(\alpha, \beta)\big)_-}$.

Note that $\rho_e^{\omega}(1,1)$ is a local assortativity measure that corresponds to the global generalized assortativity coefficient proposed in \citep{U.Pigorsch.2022}. Moreover, $\rho_e^{\omega}(0,1)$ is a local assortativity measure corresponding to the (global) assortativity coefficient suggested by \cite{Leung.2007}. These newly proposed generalized local assortativity measures allow to assess the local assortativity structure of weighted networks, which was not possible before.

So far, we have considered undirected networks. In the following we extend the proposed local assortativity measures to directed networks.

\subsection{Directed Weighted Networks}
\cite{Newman.2003} defines assortativity in directed networks as the correlation coefficient between the excess out-degree of the vertex that the $i$-th edge leads out of and the excess in-degree of the vertex that the $i$-th edge leads into.
\cite{Piraveenan.2012} additionally consider both the correlation between (excess) out-degrees of both ends of an edge and the correlation between (excess) in-degrees of both ends of an edge resulting in out-assortativity, which is the tendency of vertices to bond with others of similar out-degree as themselves and in-assortivity, which is the tendency of vertices to bond with others of similar in-degree as themselves. Overall there are four different variants of the assortativity coefficient in directed networks which we refer to as the \textit{mode} of assoartativity, namely: \textit{out-in}, \textit{out-out}, \textit{in-in} and \textit{in-out}. 

In the following, we extend these definitions to weighted networks by considering the corresponding correlation of the excess out- or in-strength instead of excess degree. To this end, let $p^{\text{out}}_{s^\prime}$ denote the \textit{out-strength} distribution, i.e., the probability that a randomly chosen vertex has out-strength $s^\prime$; and $p^{\text{in}}_{s^\prime}$ the \textit{in-strength} distribution, i.e., the probability that a randomly chosen vertex has in-strength $s^\prime$.

For the definitions of the \textit{excess out-} or \textit{in-strength} distributions we have to distinguish whether an edge \textit{leads out of} a vertex or whether it \textit{leads into} a vertex. We indicate the latter by a superscript asterisk symbol. Accordingly, $q_s^{\text{out}}$ denotes the excess out-strength distribution of the end that a directed edge \textit{leads out of}
and $q^{\ast\text{out}}_s$ the excess out-strength distribution of the end that a directed edge \textit{leads into}. Note that $q_s^{\text{out}}$ depends on the edge weight $\omega$, i.e., $s = s^\prime - \omega$, hence, $q_s^{\text{out}}$ is distributed according to $(s+\omega)p^{\text{out}}_{(s+\omega)}$ with normalized distribution:
\begin{align*}
	q^{\text{out}}_s = \frac{(s+\omega)p^{\text{out}}_{(s+\omega)}}{\sum_{s^\prime} s^\prime p^{\text{out}}_{s^\prime}}.
\end{align*}
As opposed to this, in case of $q^{\ast\text{out}}_s$ the excess out-strength $s$ of an end does not depend on the edge weight $\omega$, and, thus, $s = s^\prime$, hence, $q_s^{\ast\text{out}}$ is distributed according to $p^{\text{out}}_{s^\prime}$.

Similarly, for the \textit{in-strength}, $q^{\ast\text{in}}_s$ denotes the excess in-strength distribution of an end that a directed edge \textit{leads into} and $q_s^{\text{in}}$ the excess in-strength distribution of an end that a directed edge \textit{leads out of}.  
This time, vice versa, $q^{\ast\text{in}}_s$ depends on the edge weight $\omega$, i.e., $s = s^\prime - \omega$, such that, $q^{\ast\text{in}}_s$ is distributed according to $(s+\omega)p^{\text{in}}_{(s+\omega)}$, where normalization yields:
\begin{align*}
	q^{\ast\text{in}}_s = \frac{(s+\omega)p^{\text{in}}_{(s+\omega)}}{\sum_{s^\prime} s^\prime p^{\text{in}}_{s^\prime}}.
\end{align*}
Contrary, $q_s^{\text{in}}$ does not depend on the edge weight $\omega$, i.e., $s = s^\prime$, such that $q_s^{\text{in}}$ is distributed according to $p^{\text{in}}_{s^\prime}$.

Furthermore, we denote by $e_{st}^{\text{out--in}}$ the joint distribution of excess out- and in-strengths, i.e., the probability that a randomly chosen directed edge \textit{leads out of} a vertex with excess out-strength $s$ and \textit{into} a vertex with excess in-strength $t$. The joint distributions for the other \textit{modes of assortativity}, which are \textit{out--out}, \textit{in--in} and \textit{in--out}, are defined accordingly. For reasons of brevity, we only discuss the derivation for the \textit{out--in} mode. The derivation for the other modes is analogous.

Based on the previous definitions, the out-in weighted assortativity coefficient is given by:
\begin{align*}
	r^\omega_{\text{out--in}} = \frac{1}{\sigma_{q^{\text{out}}}\sigma_{q^{\ast\text{in}}}} \sum_{st} st (e_{st}^{\text{out--in}} - q_s^{\text{out}}q^{\ast\text{in}}_t).\numberthis\label{eq_r_omega_dir}
\end{align*}
Apparently, this is a directed version of the weighted assortativity coefficient in \cref{eq_r_omega}, and is the correlation between the excess out-strength of the outgoing end and the excess in-strength of the incoming end of an edge.

Estimating the theoretical quantities by their weighted sample counterparts, and introducing $(\alpha, \beta)$ allows to define the \textit{generalized directed edge assortativeness values} $\rho_e^{\omega}(\alpha,\beta,\text{mode})$ for any mode, e.g. for the out-in mode we obtain:
\begin{multline*}
	\rho_e^{\omega}(\alpha,\beta,\text{out--in}) =\\[5pt] \frac{\omega_e^\beta\big[l^{\text{out}}_e-\bar{U}^{\omega}_{q^{\text{out}}_s}(\alpha, \beta)\big]\big[m^{\text{in}}_e-\bar{U}^{\omega}_{q^{\ast\text{in}}_s}(\alpha, \beta)\big]}{\Omega\cdot\hat{\sigma}^{\omega}_{q^{\text{out}}_s}(\alpha, \beta)\cdot\hat{\sigma}^{\omega}_{q^{\ast\text{in}}_s}(\alpha, \beta)},\numberthis\label{eq_rho_directed_weighted_alpha_beta}
\end{multline*}
where $l^{\text{out}}_e$ and $m^{\text{in}}_e$ are similarly defined, as before, with the addition that the direction of edge $e$ is incorporated, i.e., if $\alpha = 1$, the quantities denote the excess out-strength of the end $l$ that edge $e$ leads out of and the excess in-strength of the end $m$ that edge $e$ leads into. If, however, the parameter $\alpha = 0$, the vertex values reduce to the excess out- and in-degree, respectively; $\bar{U}^{\omega}_{q^{\text{out}}_s}(\alpha, \beta) = \Omega^{-1}\sum_{e=1}^M \omega_e^\beta l_e^{\text{out}}$ and $\bar{U}^{\omega}_{q^{\ast\text{in}}_s}(\alpha, \beta) = \Omega^{-1}\sum_{e=1}^M \omega_e^\beta m_e^{\text{in}}$ are the (weight\-ed) sample mean out- and in\-/strengths of the outgoing and incoming ends, respectively; and the respective (weighted) sample standard deviations are given by $\hat{\sigma}^{\omega}_{q^{\text{out}}_s}(\alpha, \beta) = \sqrt{\Omega^{-1}\sum_{e=1}^M \omega_e^\beta \big(l_e^{\text{out}}\big)^2 - \big[\bar{U}^{\omega}_{q^{\text{out}}_s}(\alpha, \beta)\big]^2}$ and $\hat{\sigma}^{\omega}_{q^{\ast\text{in}}_s}(\alpha, \beta) = \sqrt{\Omega^{-1}\sum_{e=1}^M \omega_e^\beta \big(m^{\text{in}}_e\big)^2 - \big[\bar{U}^{\omega}_{q^{\ast\text{in}}_s}(\alpha, \beta)\big]^2}$.\footnote{Unlike in undirected networks, in a directed network each edge has one outgoing and one incoming end. Therefore, the mean excess out-degree or out-strength of an end of an edge is obtained by averaging the excess out-degrees or -strengths of the outgoing ends over all edges in the network. The mean excess in-degree or -strength of an end of an edge is obtained, accordingly. This also applies to the computation of the standard deviation of the excess out- or in-degree or -strength of an end of an edge.}

For $\beta = 1$ the weighted sample mean and standard deviation of the vertex value of an end of an edge are used, whereas for $\beta = 0$, the unweighted mean and standard deviation are employed. The expression in \cref{eq_rho_directed_weighted_alpha_beta} is the most general version of the (weighted) assortativeness values for a directed network.\footnote{Technically, \cref{eq_rho_directed_weighted_alpha_beta} is capable of handling undirected networks as well, however, a slight modification to the network is necessary, i.e., replacing each undirected edge by two directed ones that point in opposite directions, alternatively \cref{eq_rho_weighted_alpha_beta} can be used, cf. \cite{Newman.2003}.} For example, consider the parameter combination $(\alpha = 0, \beta = 0)$, the \textit{summation of the generalized directed edge assortativeness values}, $\rho_e^{\omega}(\alpha,\beta,\text{mode})$, for the \textit{out--in} mode of assortativity corresponds to the directed assortativity coefficient, $r_d$, of \cite{Newman.2003}, whereas for the \textit{out--out} and \textit{in--in} modes it corresponds to the so-called \textit{out-assortativity}, $r_{\text{out}}$, and \textit{in-assortativity}, $r_{\text{in}}$, respectively, see \cite{Piraveenan.2012}.

Moreover, in directed networks, the vertex assortativeness values can be further differentiated. The sum can either be computed using the edge assortativeness values of the outgoing edges or the incoming edges. This results in two representations of vertex assortativeness values, which we refer to as the \textit{generalized vertex out-assortativeness values} and \textit{generalized vertex in-assortativeness values}, respectively. The generalized vertex out-assortativeness and in-assortativeness values of a vertex $v$ are denoted by $\rho^\omega_v(\alpha, \beta, \textit{out})$ and $\rho^\omega_v(\alpha, \beta, \textit{in})$, respectively, and they are defined as:
\begin{align*}
	\rho^\omega_v(\alpha, \beta, \textit{out}) = \sum_{u=1}^n \rho^\omega_{e_{vu}}(\alpha, \beta, \text{mode}),\\[5pt]
	\rho^\omega_v(\alpha, \beta, \textit{in}) = \sum_{u=1}^n \rho^\omega_{e_{uv}}(\alpha, \beta, \text{mode}).
\end{align*}

We suggest that the directed vertex assortativeness values are chosen in accordance with the considered mode of assortativity. For example, both generalized vertex out- and in-assortativeness values can be determined for the \textit{out--in} mode of assortativity. For the generalized vertex out-assortativeness, the contribution to the global assortativity is assigned to the vertex from which the edge originates. In the case of generalized vertex in-assortativeness values, the contribution to global assortativity is assigned to the vertex pointed to by an edge. It depends on the research setting, in particular which vertex is considered to be responsible for a connection, whether the former or the latter should be preferred. Finally, for the generalized vertex out- and in-assortativeness values it holds that:
\begin{align*}
	\sum_{v=1}^n \rho^\omega_v(\alpha, \beta, \textit{out})
    &= \sum_{v=1}^n \rho^\omega_v(\alpha, \beta, \textit{in})\\
    &= \sum_{uv} \rho^\omega_{e_{uv}}(\alpha, \beta, \text{mode})\\
    &= r^\omega(\alpha, \beta, \text{mode}).\numberthis\label{eq_generalized_vertex_assortativeness}
\end{align*}

To summarize, in this section we have generalized the concept of local assortativity by introducing the parameters $\alpha$ and $\beta$. As a result, it is now possible to analyze local assortativity also in weighted networks and not only in unweighted ones. Furthermore, in directed networks, local assortativity can be assessed for the four different modes.
 
\section{Empirical Analysis of Local Assortativity Patterns}\label{sec_empirical_local_assortativity_analysis}

In the following, we demonstrate the usefulness of the proposed generalized local assortativity measures by presenting an in-depth analysis of the generalized local assortativity of theoretical network models as well as of real world directed and undirected weighted networks. 

\subsection{Selected Network Models}\label{sec_network_models}
We consider weighted extensions of the \textit{Erd\H{o}s-Rényi random graph} (ER) and Barabási-Albert (BA) models. Both models are known to be (globally) non-assortative, and we expect this to hold for their weighted extensions as well, especially since the weighted versions nest the unweighted ones. A main purpose of the analysis is to show that our generalized approach to measuring assortativity is able to identify a known (global) non-assortative network and to reveal potential differences in the local assortativity structure. In particular, we show below that both considered networks are globally non-assortative, but differ in their local assortativity structure. This illustrates the usefulness of our generalized local assortativity measures, as they allow further differentiation of the topology of networks with similar global assortativity.

\subsubsection{The Weighted Random Graph Model}

The \textit{weighted random graph} (WRG) model of \cite{Garlaschelli.2009b} is an extension of the \textit{Erd\H{o}s-Rényi random graph} (ER) model towards weighted networks. It seizes on the $G_{n,p}$ ensemble, cf. \cite{Gilbert.1959}, but incorporates edge weights $\omega$. The derivation is analogous to the ER model, see \cite{Park.2003, Maslov.2004, Garlaschelli.2008, Garlaschelli.2009} and \cite{Garlaschelli.2009b} for a thorough derivation of the model, and \cite{Garlaschelli.2009b} and \cite{Coolen.2017} on how to draw samples from it.
In essence, the WRG model is completely specified by the probability that any two vertices $i$ and $j$ are connected by an edge of weight $\omega$, which is given by:
\begin{align*}
	P(W_{ij}=\omega) = P(\omega) = p^\omega(1-p).\numberthis\label{eq_wrg_prob}
\end{align*}
The probability that there is no edge between two vertices is denoted by $P(0)$, and, thus, $1- P(0) = p$ is the probability of two vertices being connected by an edge of any (non-zero) weight. 

\begin{table*}[!htbp]
	\centering
	\scriptsize
	\renewcommand{\arraystretch}{1.1}
	\begin{tabular*}{\textwidth}{l @{\extracolsep{\fill}} rrrr}
		\toprule
		\multirow{3}[4]{*}{Measure}&\multicolumn{4}{c}{$\bar{\omega} = 0.02$}\\
		\cmidrule{2-5}
		& \multicolumn{2}{c}{$\alpha = 0$} & \multicolumn{2}{c}{$\alpha = 1$}\\
		\cmidrule(r){2-3}\cmidrule(l){4-5} 
		&
		\multicolumn{1}{c}{$\beta = 0$} & \multicolumn{1}{c}{$\beta = 1$} &
		\multicolumn{1}{c}{$\beta = 0$} & \multicolumn{1}{c}{$\beta = 1$} \\
		\midrule
		\multicolumn{5}{l}{undirected} \\
		\midrule
		$r^\omega$&-0.002&-0.002&-0.002&-0.002 \\ 
		$P\big(\rho^\omega_e > 0\big)$& 0.500& 0.500& 0.502& 0.501 \\ 
		$\overline{\big(\rho^\omega_e\big)_+}$& 6.48e-05& 6.48e-05& 6.43e-05& 6.44e-05 \\ 
		$\overline{\big(\rho^\omega_e\big)_-}$& 6.53e-05& 6.53e-05& 6.52e-05& 6.53e-05 \\ 
		$P\big(\rho^\omega_v > 0\big)$& 0.499& 0.499& 0.499& 0.499 \\
		\midrule
		\multicolumn{5}{l}{out--in} \\
		\midrule
		$r^\omega$&-0.001&-0.001&-0.001&-0.001 \\ 
		$P\big(\rho^\omega_e > 0\big)$& 0.500& 0.500& 0.501& 0.501 \\ 
		$\overline{\big(\rho^\omega_e\big)_+}$& 3.26e-05& 3.26e-05& 3.23e-05& 3.23e-05 \\ 
		$\overline{\big(\rho^\omega_e\big)_-}$& 3.27e-05& 3.27e-05& 3.25e-05& 3.25e-05 \\ 
		$P\big(\rho^\omega_v(\text{out}) > 0\big)$& 0.499& 0.499& 0.500& 0.500 \\ 
		$P\big(\rho^\omega_v(\text{in}) > 0\big)$& 0.500& 0.500& 0.500& 0.501 \\
		\midrule
		\multicolumn{5}{l}{out--out} \\
		\midrule
		$r^\omega$&-4.25e-04&-2.63e-04&-4.09e-04&-2.46e-04 \\ 
		$P\big(\rho^\omega_e > 0\big)$& 0.500& 0.500& 0.501& 0.501 \\ 
		$\overline{\big(\rho^\omega_e\big)_+}$& 3.26e-05& 3.26e-05& 3.23e-05& 3.23e-05 \\ 
		$\overline{\big(\rho^\omega_e\big)_-}$& 3.27e-05& 3.27e-05& 3.25e-05& 3.25e-05 \\ 
		$P\big(\rho^\omega_v(\text{out}) > 0\big)$& 0.499& 0.500& 0.498& 0.499 \\ 
		$P\big(\rho^\omega_v(\text{in}) > 0\big)$& 0.500& 0.501& 0.500& 0.500 \\
		\midrule
		\multicolumn{5}{l}{in--in} \\
		\midrule
		$r^\omega$&-0.001&-0.001&-0.001&-0.001 \\ 
		$P\big(\rho^\omega_e > 0\big)$& 0.500& 0.500& 0.501& 0.501 \\ 
		$\overline{\big(\rho^\omega_e\big)_+}$& 3.25e-05& 3.25e-05& 3.22e-05& 3.22e-05 \\ 
		$\overline{\big(\rho^\omega_e\big)_-}$& 3.27e-05& 3.27e-05& 3.25e-05& 3.25e-05 \\ 
		$P\big(\rho^\omega_v(\text{out}) > 0\big)$& 0.499& 0.498& 0.499& 0.499 \\ 
		$P\big(\rho^\omega_v(\text{in}) > 0\big)$& 0.497& 0.499& 0.499& 0.500 \\
		\midrule
		\multicolumn{5}{l}{in--out} \\
		\midrule
		$r^\omega$&-4.49e-04&-0.001&-4.05e-04&-4.77e-04 \\ 
		$P\big(\rho^\omega_e > 0\big)$& 0.501& 0.501& 0.503& 0.503 \\ 
		$\overline{\big(\rho^\omega_e\big)_+}$& 3.26e-05& 3.26e-05& 3.22e-05& 3.22e-05 \\ 
		$\overline{\big(\rho^\omega_e\big)_-}$& 3.27e-05& 3.27e-05& 3.26e-05& 3.26e-05 \\ 
		$P\big(\rho^\omega_v(\text{out}) > 0\big)$& 0.499& 0.499& 0.499& 0.499 \\ 
		$P\big(\rho^\omega_v(\text{in}) > 0\big)$& 0.499& 0.499& 0.499& 0.499 \\ 	
		\bottomrule
		\end{tabular*}
		\caption[Generalized local assortativity analysis of the WRG model for different parameter combinations $(\alpha,\beta)$.]{\textbf{Generalized local assortativity analysis of the WRG model for different parameter combinations $(\alpha,\beta)$.} Reported are the generalized assortativity coefficient $r^\omega$, fraction of local assortative edges \mbox{$P(\rho^\omega_e > 0)$}, average absolute magnitude of assortative edges $\overline{(\rho^\omega_e)_+}$, average absolute magnitude of disassortative edges $\overline{(\rho^\omega_e)_-}$, fraction of local assortative vertices $P(\rho^\omega_v > 0)$ (undirected), and fraction of local out- and in-assortative vertices $P(\rho^\omega_v(\text{out}) > 0)$ and $P(\rho^\omega_v(\text{in}) > 0)$ (directed), respectively, for the WRG model, with a target mean edge weight $\bar{\omega} = 0.02$. An ensemble of $100$ weighted random graphs (WRG) of order $n = 1000$ is drawn for each mode of assortativity. The results are averaged over the samples of the ensemble.
 		}
 		\label{tab_wrg_local_assortativity_results}
\end{table*}

\begin{figure*}[!htbp]
	\centering
	\begin{subfigure}{.45\textwidth}
		\centering
		\caption{}
		\includegraphics[width = \textwidth]{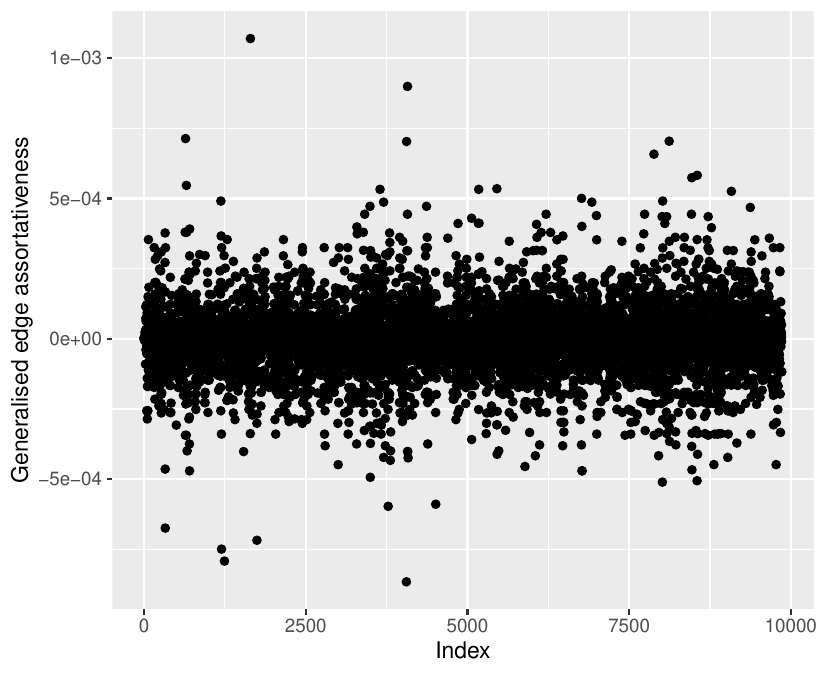}
		\label{wrg_edge_assortativeness_index}
	\end{subfigure}
	\hfill
	\begin{subfigure}{.45\textwidth}
		\centering
		\caption{}
		\includegraphics[width = \textwidth]{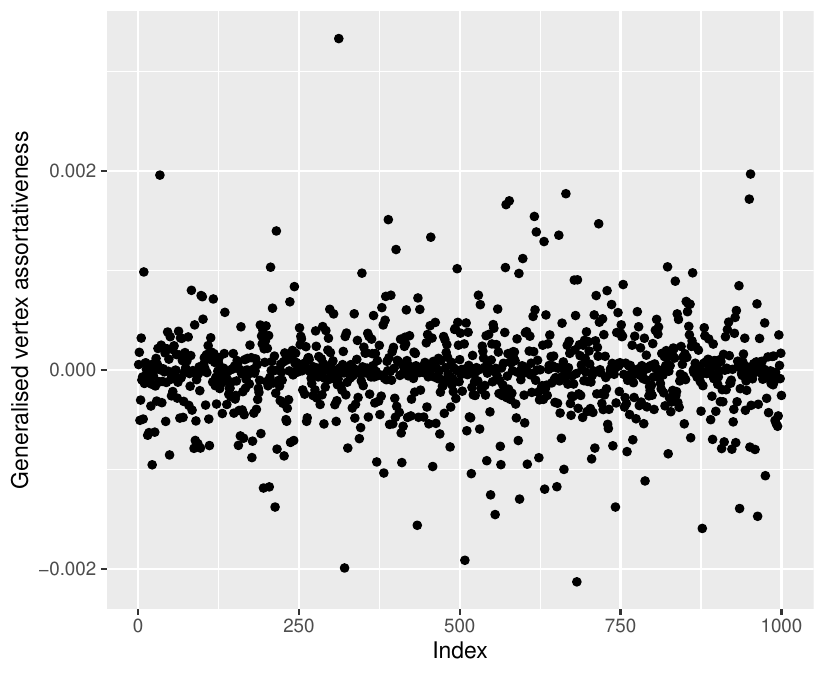}
		\label{wrg_vertex_assortativeness_index}
	\end{subfigure}
	\hfill
	\begin{subfigure}{.45\textwidth}
		\centering
		\caption{}
		\includegraphics[width = \textwidth]{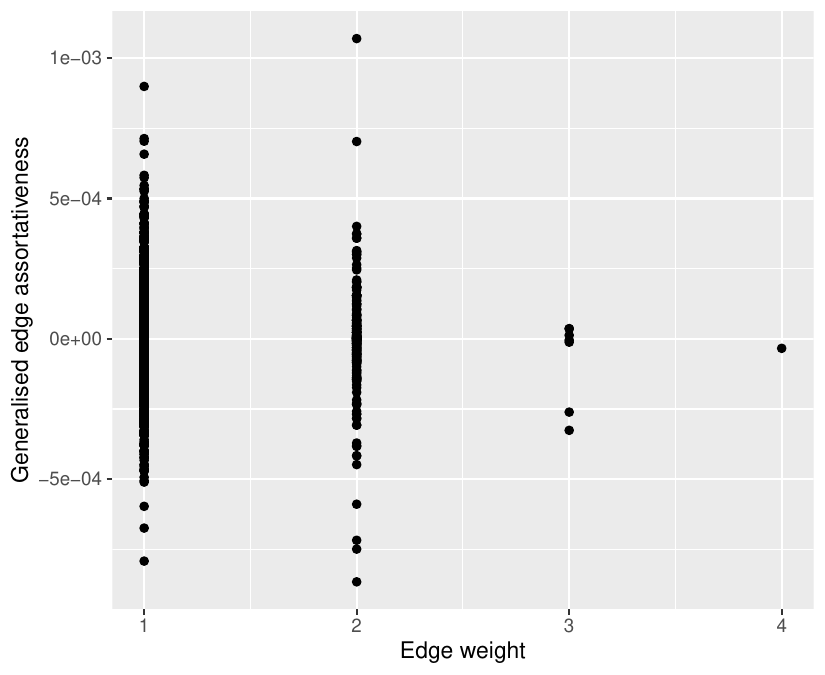}
		\label{wrg_edge_assortativeness_edge_weight}
	\end{subfigure}
	\hfill
	\begin{subfigure}{.45\textwidth}
		\centering
		\caption{}
		\includegraphics[width = \textwidth]{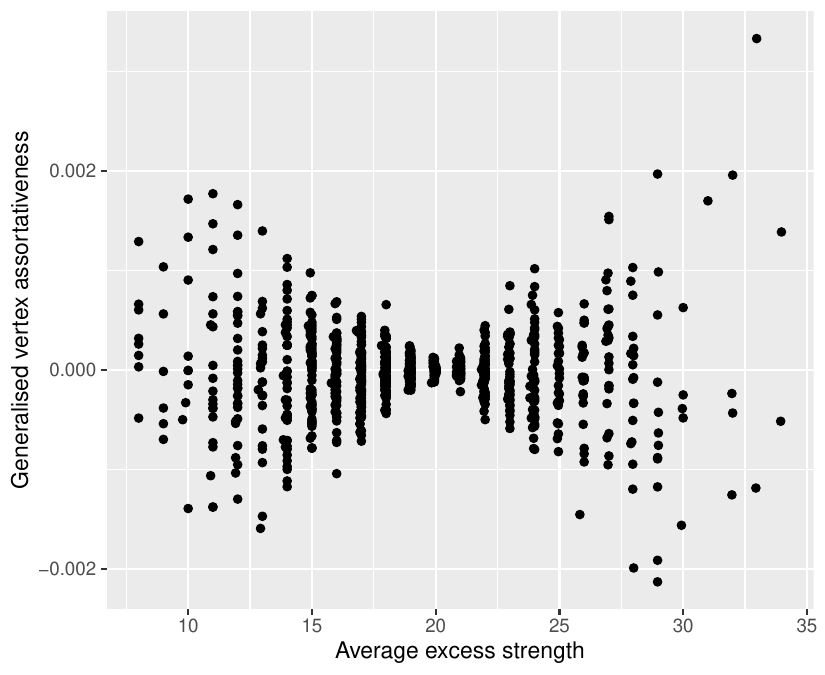}
		\label{wrg_vertex_assortativeness_excess_strength}
	\end{subfigure}
	\hfill
	\begin{subfigure}{.45\textwidth}
		\centering
		\caption{}
		\includegraphics[width = \textwidth]{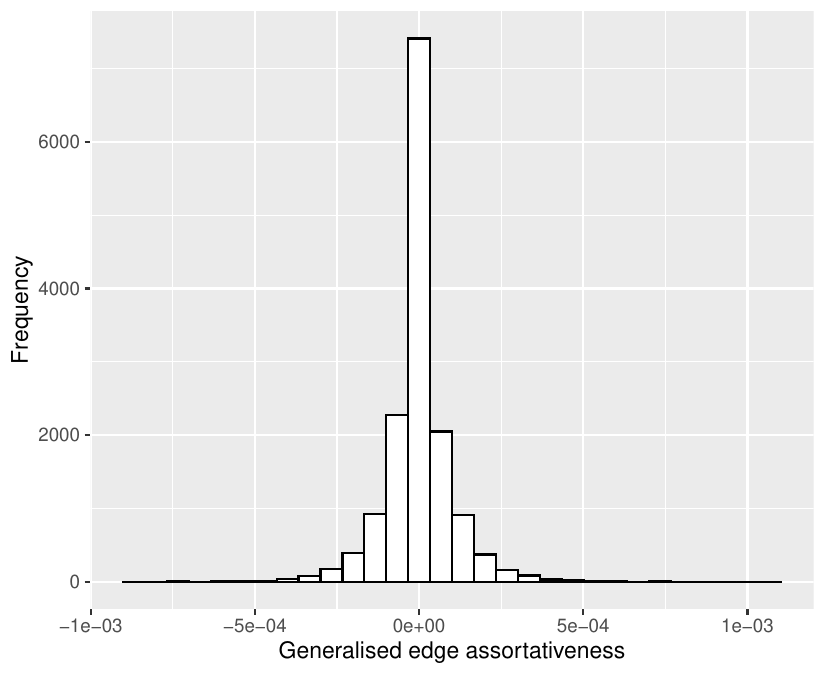}
		\label{wrg_edge_assortativeness_histogram}
	\end{subfigure}
	\hfill
	\begin{subfigure}{.45\textwidth}
		\centering
		\caption{}
		\includegraphics[width = \textwidth]{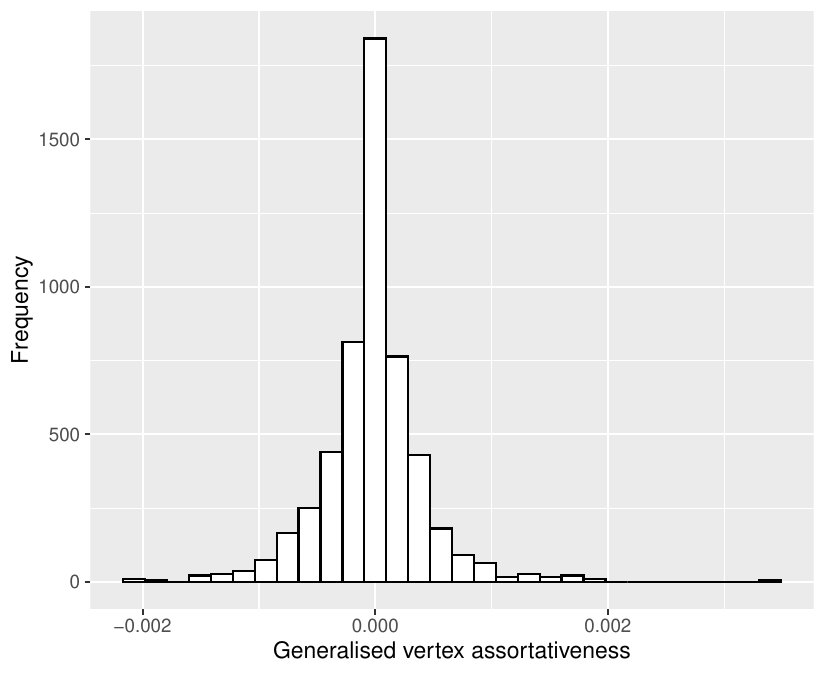}
		\label{wrg_vertex_assortativeness_histogram}
	\end{subfigure}
	\caption[Generalized local assortativity plots of the WRG model.]{\textbf{Generalized local assortativity plots of the WRG model.} The plots are computed for a single undirected sample of the WRG model of order $n=1000$ with an average edge weight $\bar{\omega}=0.02$, for the parameter combination $(\alpha=1, \beta=1)$. Depicted are scatter plots of the generalized edge assortativeness values, $\rho_e^\omega(\alpha, \beta)$ against the edge index (a), of the generalized vertex assortativeness values, $\rho_v^\omega(\alpha, \beta)$, against the vertex index (b), of $\rho_e^\omega(\alpha, \beta)$ against the edge weight (c), of $\rho_v^\omega(\alpha, \beta)$ against average excess strength (d), as well as the histograms of $\rho_e^\omega(\alpha, \beta)$ and $\rho_v^\omega(\alpha, \beta)$, (e) and (f), respectively.} 
	\label{fig:wrg_sample_plots}
\end{figure*}

Since most real-world networks are supercritical, cf. \cite{Barabasi.2016}, we base our local assortativity analysis of the WRG model on that regime. For each mode of assortativity, i.e., \textit{undirected}, \textit{out--in}, \textit{out--out}, \textit{in--in} and \textit{in--out}, we sample an ensemble of $N=100$ networks. We set the average edge weight of a sample to $\bar{\omega} = \frac{p}{1-p} = 0.02$. This way we ensure that the resulting samples are in the supercritical regime. The range of $p$ for which the supercritical regime results is $\frac{1}{N} < p < \frac{\operatorname{ln}N}{N}$, cf \cite[p. 86]{Barabasi.2016}. An average edge weight of $\bar{\omega} = 0.02$ corresponds to a connection probability $p = 0.0196$, which lies in between $\frac{1}{100} = 0.01 < 0.0196 < 0.0461 = \frac{\operatorname{ln}100}{100}$. Any other value from this range would be appropriate, too. 

For each of these simulated networks we compute the generalized local assortativity measures. \Cref{tab_wrg_local_assortativity_results} presents the averaged results (with averages taken over the samples of the ensemble) for the four different parameter combinations $(\alpha, \beta) \in \{0,1\}$. Note that, for the parameter combination $(\alpha = 0, \beta = 0)$, the WRG model is projected onto an unweighted graph, i.e., analyzing the generalized assortativity of the WRG model is the same as analyzing the assortativity of the ER model, for this combination. The following detailed description refers to all modes and parameter combinations, as the results are quite similar. For example, the generalized assortativity coefficient $r^\omega(\alpha, \beta)$ equals almost zero, indicating that the WRG model generates networks that do not show any degree or strength correlations, as expected.\footnote{The fact that the coefficient is not exactly zero might be explained by structural constraints, i.e., the finite size of the network sample, cf. \cite{Serrano.2006} and \cite{Yang.2017}.} Moreover, there is an equal proportion of assortative and disassortative edges, as $P(\rho_e>0) \approx 0.5$. Also, assortative and disassortative edges tend to be equally strong as the average absolute magnitude of assortative and disassortative edges is fairly balanced. Finally, there is an equal proportion of assortative and disassortative vertices, as $P(\rho_v>0) \approx 0.5$.  

\Cref{wrg_edge_assortativeness_index,wrg_vertex_assortativeness_index,wrg_edge_assortativeness_edge_weight,wrg_vertex_assortativeness_excess_strength,wrg_edge_assortativeness_histogram,wrg_vertex_assortativeness_histogram} provide a graphical analysis of the local assortativity pattern of the undirected WRG model, for the parameter combination $(\alpha = 1, \beta = 1)$.\footnote{We forgo showing figures for the other parameter combinations and modes of assortativity as they look similar. They are available upon request.} More precisely, \Cref{wrg_edge_assortativeness_index} shows a scatter plot of the relation between the edge assortativeness values and the position of an edge in the edge list or adjacency matrix. 
If the elements of the edge list or adjacency matrix of the network are ordered from a content perspective, this plot will provide a tool for analyzing the edge assortativeness with respect to this ordering. In the case of the theoretical WRG, however, there is no particular ordering of vertices or edges. Therefore, \Cref{wrg_edge_assortativeness_index,wrg_vertex_assortativeness_index} emphasize once more that no degree or strength correlations are present in this model, as positive as well as negative edge and vertex assortativeness values are evenly distributed among edges and vertices, respectively. This observation is also supported by the histograms in \Cref{wrg_edge_assortativeness_histogram,wrg_vertex_assortativeness_histogram}. However, \Cref{wrg_vertex_assortativeness_excess_strength} shows an interesting pattern. Although vertices tend to be locally non-assortative on average, the figure shows that there is more variation in the vertex assortativeness values for vertices with an average excess strength either below or above a particular value.\footnote{The average excess strength of a vertex corresponds to the total strength less the mean of the weights of the edges emanating from that particular vertex.
} The value for which the variation of generalized vertex values is lowest $(\approx 20)$ happens to be the global average excess strength of the ends of an edge. This, however, is due to the way how (edge-based) local assortativity is defined, as the definition pivots on the global average excess strength, $\bar{U}_{q_s}$, cf. \cref{eq_third}.

A main purpose of the preceding analysis is verifying that our generalized local assortativeness measures are able to identify a known non-assortative network. It is, thus, pleasant to see that the results coincide with our expectations: the WRG model shows neither global nor local assortative or disassortative tendencies, just as its unweighted counterpart the ER model with unweighted assortativity.

We continue with another theoretical model that, similar to the WRG, is known to be (globally) non-assortative, but is structurally different with respect to local assortativity. We show that these different structures can be uncovered by our local assortativity measures.

\subsubsection{Weighted Preferential Attachment Models}

\cite{Yook.2001,Zheng.2003} consider weighted scale free networks.
The model by \cite{Yook.2001} is referred to as the \textit{weighted scale free model} (WSF) and expands the simple BA model by \cite{Barabasi.1999} by a weight assignment scheme, i.e., edges are created according to the simple preferential attachment scheme in \cite{Barabasi.1999}, to which weights are assigned afterwards. \cite{Zheng.2003} contribute to this by altering the weight assignment scheme of the WSF model to a \textit{stochastic weight assignment scheme}.
At first, random fitness values $\eta$ are assigned to the vertices, where, for simplicity, $\eta_i \sim U_{[0,1]}$ for each vertex $i$.\footnote{However, the fitness values are not restricted to be uniformly distributed, but rather can be distributed according to any distribution $P(\eta)$, cf. \cite{Zheng.2003}.} With probability $p$, edge weights are then assigned to newly created edges according to the \textit{connectivity driven weight} scheme suggested by \cite{Yook.2001}. With probability $(1-p)$ edge weights are assigned according to the scheme:
\begin{align}
	w_{ji} = \frac{\eta_i}{\sum_{\{i^\prime\}} \eta_{\{i^\prime\}}}.
	\label{eq_fitness_weight_assignment_scheme}
\end{align}
For $p = 1$ the scheme by \cite{Yook.2001} is recovered, whereas, for $p = 0$ weights are driven entirely by vertex fitness. For values in the range $0<p<1$, weights are randomly assigned, mimicking the real-world behaviour of newcomers entering a network, and choosing to connect to others either based on their popularity or based on other non-popularity related attributes captured by the fitness value. As with the simple BA model, there exists a growing number of extensions to the weighted BA model, see for example \cite{Coolen.2017,Barthelemy.2005}. An extension towards directed networks has been introduced by \cite{Bollobas.2003}.

To the best of our knowledge, so far, there is no established model for producing random \textit{directed and weighted} scale free networks. An attempt at a definition of such a model has been made by \cite{Yuan.2021} but fails to generalize the model by \cite{Bollobas.2003} completely.\footnote{The case in which two randomly chosen existing vertices are connected is not contained in the model of \cite{Yuan.2021}.} Moreover, their choice of assigning edge weights in the process of link formation seems a bit arbitrary to us, as they randomly sample integer values ranging from $1$ to $10$.\footnote{We aim at analyzing the influence of edge weights on the local (edge) assortativity, and thus, it does not seem sensible to us to set edge weights arbitrarily and then to interpret them.} We conclude that this model needs further investigation, and therefore, only consider the undirected WSF model in this paper. We postpone the analysis of directed weighted scale-free models to future research.

For our assortativity analysis, we consider the WSF model with stochastic weights assignment scheme, for which we set the initial number of vertices $m_0 = 5$, the number of steps $T = 10000$, the number of newly created edges per step $m = 2$, and the connection probability $p = 0.5$, resulting in a representative of a weighted network with a scale free degree (and strength) distribution. \Cref{tab_wsf_analysis} presents the results of the local assortativity analysis for the WSF model. Reported is the average of each (local) assortativity measure over the individual samples of the ensemble. Interestingly, the results show that the model generates predominantly disassortative networks, as the generalized assortativity coefficient, $r^\omega$, is negative for all parameter combinations $(\alpha, \beta)$. This is the result of the fact that, on the one hand, the proportion of disassortative edges is slightly higher, i.e., $P(\rho_e^\omega > 0) > 0.5$, and, on the other hand, the average absolute magnitude of disassortative edges is higher than that of assortative edges, i.e., $\overline{\big(\rho^\omega_e\big)_-} > \overline{\big(\rho^\omega_e\big)_+}$. Furthermore, most of the nodes are disassortative, since $P\big(\rho^\omega_v > 0\big) > 0.5$.

\begin{table*}[!htb]
	\centering
	\scriptsize
	\begin{tabular*}{\textwidth}{l @{\extracolsep{\fill}} rrrr}
		\toprule
		\multirow{2}[2]{*}{Measure}&
		\multicolumn{2}{c}{$\alpha = 0$} & \multicolumn{2}{c}{$\alpha = 1$}\\
		\cmidrule(r){2-3}\cmidrule(l){4-5} 
		&
		\multicolumn{1}{c}{$\beta = 0$} & \multicolumn{1}{c}{$\beta = 1$} &
		\multicolumn{1}{c}{$\beta = 0$} & \multicolumn{1}{c}{$\beta = 1$} \\
		\midrule
		$r^\omega$&-0.042&-0.079&-0.039&-0.074 \\ 
		$P\big(\rho^\omega_e > 0\big)$& 0.671& 0.718& 0.706& 0.748 \\ 
		$\overline{\big(\rho^\omega_e\big)_+}$& 6.48e-06& 5.57e-06& 5.53e-06& 4.87e-06 \\ 
		$\overline{\big(\rho^\omega_e\big)_-}$& 1.95e-05& 2.81e-05& 1.98e-05& 2.91e-05 \\ 
		$P\big(\rho^\omega_v > 0\big)$& 0.692& 0.702& 0.714& 0.725 \\ 
		\bottomrule
		\end{tabular*}
		\caption[Generalized assortativity analysis of the WSF model for different parameter combinations $(\alpha,\beta)$.]{\textbf{Generalized assortativity analysis of the WSF model for different parameter combinations $(\alpha,\beta)$.} Reported are the generalized assortativity coefficient $r^\omega$, fraction of local assortative edges $P(\rho^\omega_e > 0)$, average absolute magnitude of assortative edges $\overline{(\rho^\omega_e)_+}$, average absolute magnitude of disassortative edges $\overline{(\rho^\omega_e)_-}$, fraction of local assortative vertices $P(\rho^\omega_v > 0)$  for the WSF model.
		An ensemble of $100$ samples of the WSF model with $m_0 = 5$, $T = 10000$, $m = 2$ and $p = 0.5$ is drawn. The results are averaged over the samples of the ensemble.}
		\label{tab_wsf_analysis}
\end{table*}

The BA model, of which the WSF model is an extension of, can be shown to be non-assortative in the limit of large $n$, cf. \cite{Newman.2002}. However, due to structural constraints such as the finite size of the network samples, the model does not produce purely non-assortative networks, cf. \cite{Maslov.2004}, \cite{Serrano.2006} and \cite{Yang.2017}. Apparently, the WSF model shares this property.

\begin{figure*}[!htbp]
	\centering
	\begin{subfigure}{.45\textwidth}
		\centering
		\caption{}
		\includegraphics[width = \textwidth]{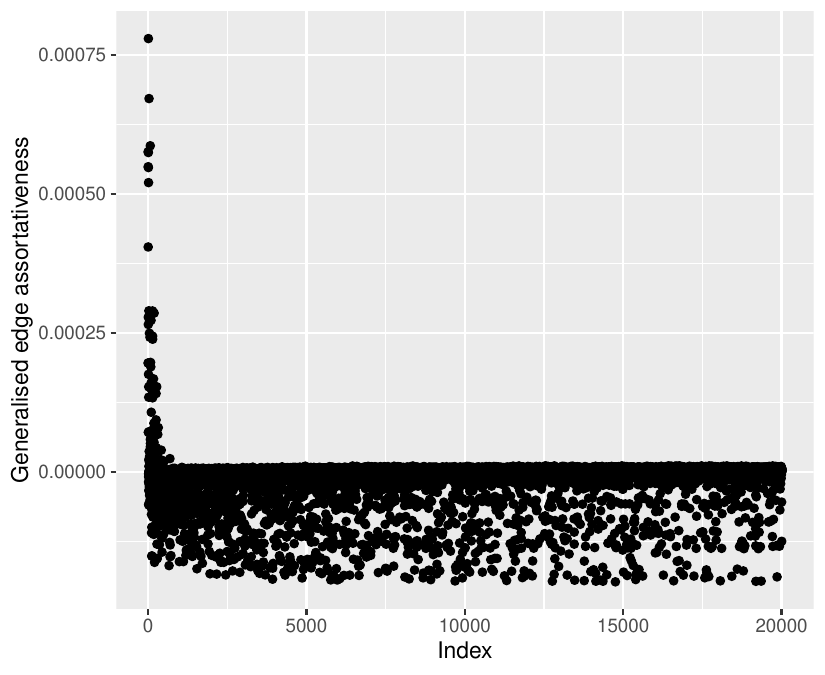}
		\label{wsf_edge_assortativeness_index}
	\end{subfigure}
	\hfill
	\begin{subfigure}{.45\textwidth}
		\centering
		\caption{}
		\includegraphics[width = \textwidth]{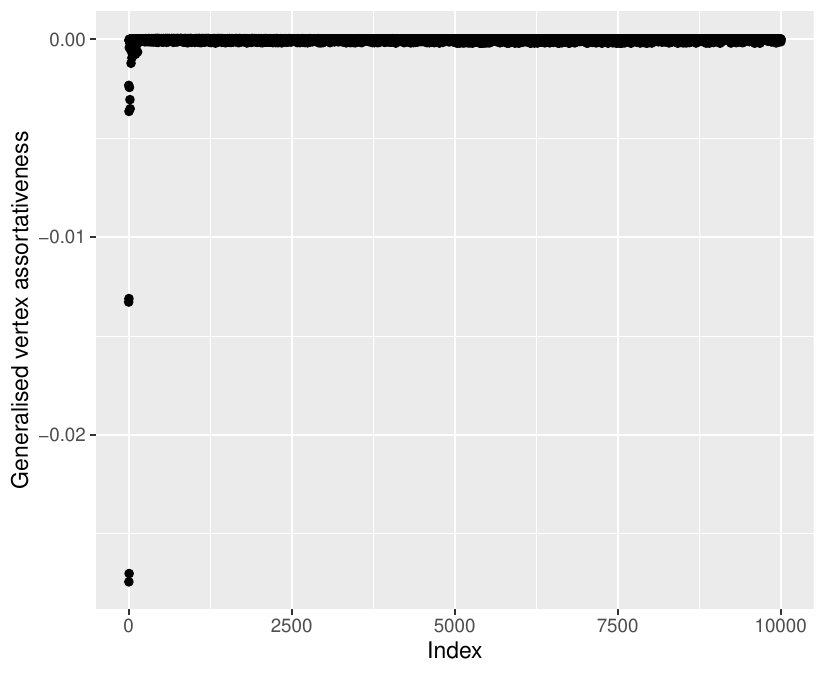}
		\label{wsf_vertex_assortativeness_index}
	\end{subfigure}
	\hfill
	\begin{subfigure}{.45\textwidth}
		\centering
		\caption{}
		\includegraphics[width = \textwidth]{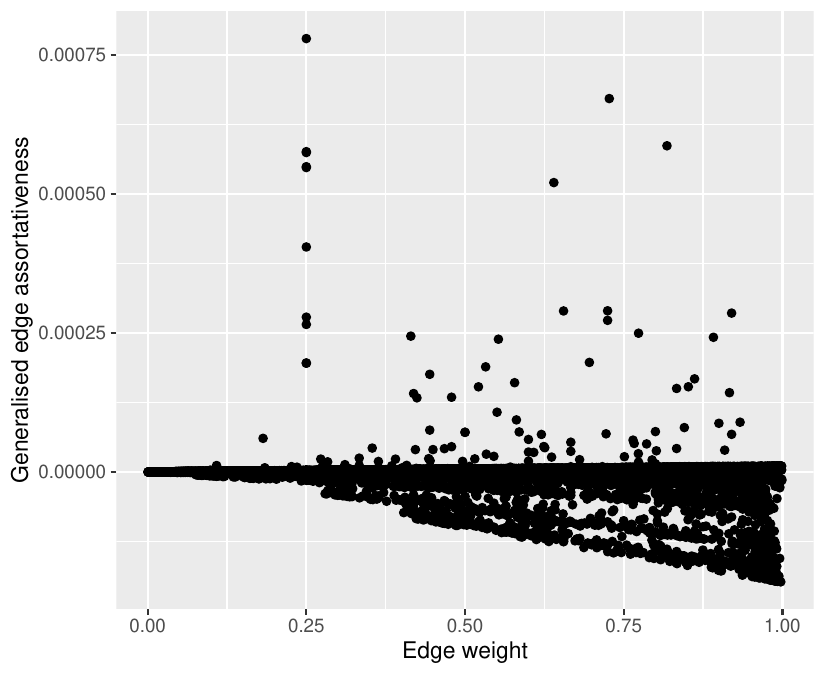}
		\label{wsf_edge_assortativeness_edge_weight}
	\end{subfigure}
	\hfill
	\begin{subfigure}{.45\textwidth}
		\centering
		\caption{}
		\includegraphics[width = \textwidth]{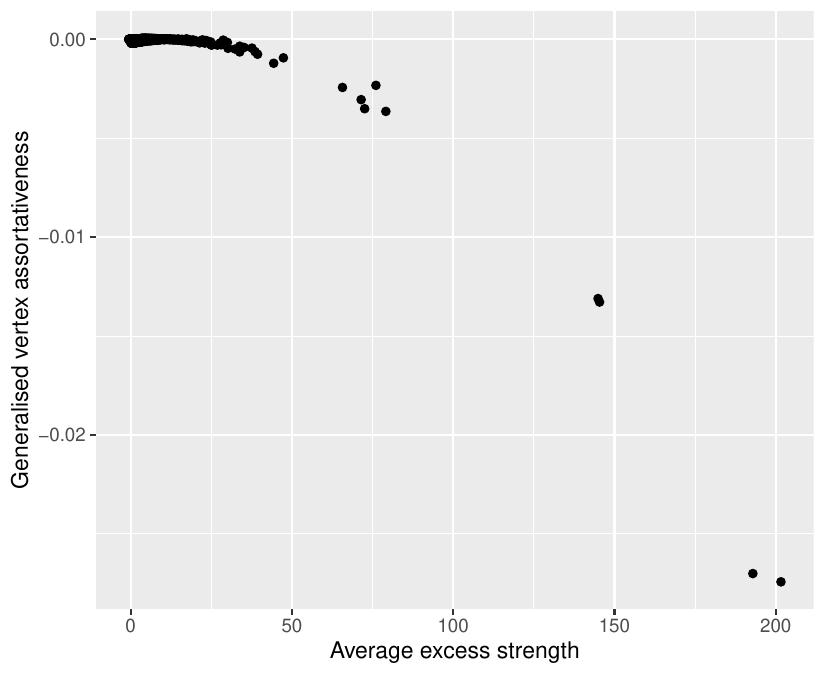}
		\label{wsf_vertex_assortativeness_excess_strength}
	\end{subfigure}
	\hfill
	\begin{subfigure}{.45\textwidth}
		\centering
		\caption{}
		\includegraphics[width = \textwidth]{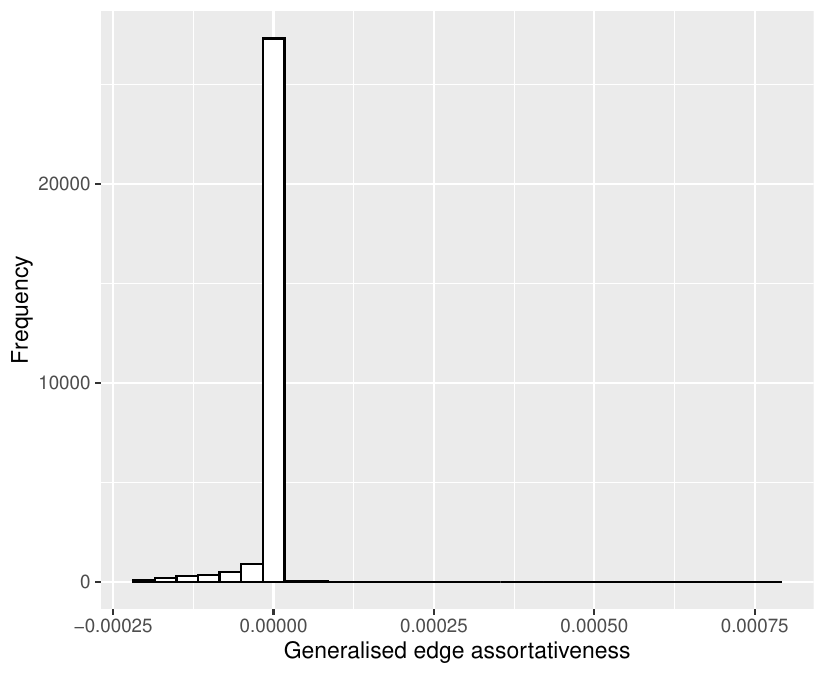}
		\label{wsf_edge_assortativeness_histogram}
	\end{subfigure}
	\hfill
	\begin{subfigure}{.45\textwidth}
		\centering
		\caption{}
		\includegraphics[width = \textwidth]{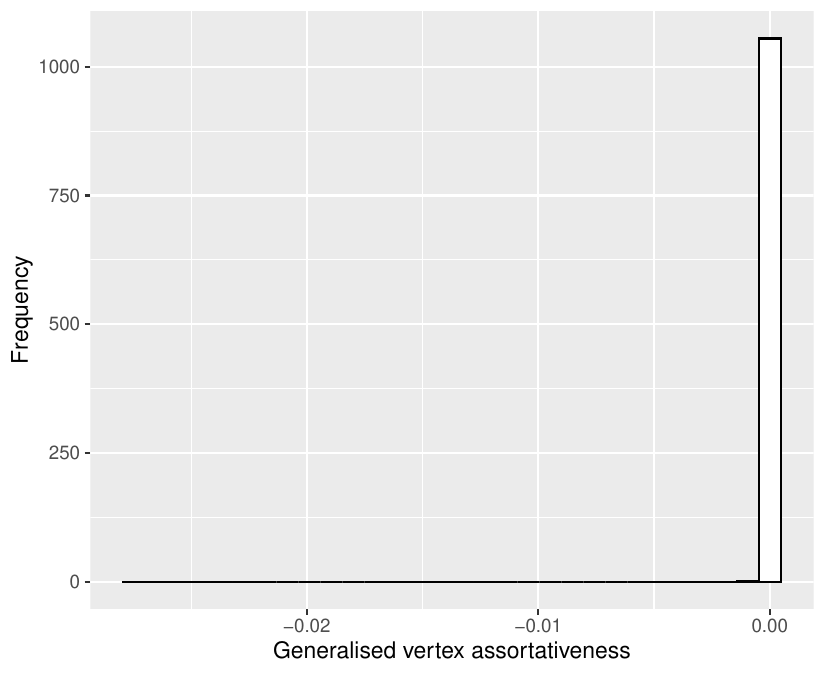}
		\label{wsf_vertex_assortativeness_histogram}
	\end{subfigure}
	\caption[Generalized local assortativity plots of the WSF model.]{\textbf{Generalized local assortativity plots of the WSF model.} The plots are computed for a single sample of the WSF model with $m_0 = 5$, $T=10000$, $m=2$ and $p=0.5$, for the parameter combination $(\alpha=1, \beta=1)$. Depicted are scatter plots of the generalized edge assortativeness values, $\rho_e^\omega(\alpha, \beta)$ against the edge index (a), of the generalized vertex assortativeness values, $\rho_v^\omega(\alpha, \beta)$, against the vertex index (b), of $\rho_e^\omega(\alpha, \beta)$ against the edge weight (c), of $\rho_v^\omega(\alpha, \beta)$ against average excess strength (d), as well as the histograms of $\rho_e^\omega(\alpha, \beta)$ and $\rho_v^\omega(\alpha, \beta)$, (e) and (f), respectively.}
	\label{fig:wsf_sample_plots}
\end{figure*}

\Cref{fig:wsf_sample_plots} presents the graphical analysis of the local assortativity of the WSF model. Obviously, in contrast to the WRG model, the WSF model exhibits some patterns in the local assortativity structure. Note that in preferential attachment models, if vertices and edges are numbered consecutively when entering the network or when forming, the respective indices correspond to the time points at which a vertex was created or at which an edge was formed. For example, in \Cref{wsf_edge_assortativeness_index,wsf_vertex_assortativeness_index} it can be seen that nodes that join the network earlier tend to be disassortative, and edges that form early tend to be assortative, cf. \cite{Noldus.2015}, who note a similar pattern in the BA model.

On the other hand, plotting the edge assortativeness values against the corresponding edge weight, cf. \Cref{wsf_edge_assortativeness_edge_weight} shows that highly weighted edges tend to be disassortative. Similarly, plotting the vertex assortativeness values against the average excess strength of a node, cf. \Cref{wsf_vertex_assortativeness_excess_strength} shows that the vertices with higher excess strength tend to be disassortative. However, the histograms in \Cref{wsf_edge_assortativeness_histogram,wsf_vertex_assortativeness_histogram} indicate that the majority of edges and vertices are non-assortative.

Moreover, most of the assortative edges can be traced back to connections between the disassortative hubs, which in turn are the product of the initialization of the model, i.e., vertices that belong to the initial set of vertices, $m_0$. Apparently, it takes some time for these initialization effects to average out, and $T = 10000$ does not seem to be sufficient for the model to adopt an overall non-assortative structure. \cite{Xu.2009} and others consider cases in which the assortativity structure of (scale free) networks is significantly influenced by super rich nodes. They argue that it may be sensible to exclude these super rich nodes from consideration when calculating certain network measures, e.g, assortativity. Given our finding, this might also be a sensible approach for the analysis of the WSF model.

Interestingly, both the WRG and the WSF model show very different local assortativity patterns, even though they are weighted extensions of models that are considered to generate global non-assortative networks. This implies that networks may exhibit a similar global assortativity, but the underlying local structures can differ substantially. Accordingly, it seems sensible to also consider the local assortativity structures of a network in order to describe its topology as precisely as possible. In the following, we turn our focus to analyzing the local assortativity patterns of two real-world networks.

\subsection{Real World Networks}\label{sec_real_world_networks}

To illustrate the analysis of generalized local assortativity of weighted real-world networks we consider the well-known undirected \textit{NetScience} network of \cite{Newman.2001b} and the commonly used directed neural network of the nematode worm \textit{Caenorhabditis elegans} (C. Elegans), cf. \cite{Watts.1998,White.1986}. The corresponding datasets are publicly available, which ensures that the subsequent results can be easily reproduced.

\subsubsection{NetScience Scientific Collaboration Network}

The NetScience network is an undirected collaboration network of scientists working on network theory, where vertices represent scientists and edges indicate if both co-authored one or more publications, cf. \cite{Newman.2001b}. The intensity of the relation between two scientists is incorporated by positive edge weights. In particular, degree in the network corresponds to the number of different co-authors of a scientist, whereas strength corresponds to the number of papers a scientist has co-authored with others, cf. \cite{Newman.2001b}.

As already noted in \cite{U.Pigorsch.2022}, the \textit{NetScience} network is an overall assortative network. In particular, the generalized global assortativity coefficients is $r^\omega_{(0,0)} = 0.4616$ (see also \Cref{tab_generalised_local_assortativity_analysis_real_networks})  indicating that scientists have a tendency to collaborate with others who are similar based on the number of co-authors (degree). Moreover, taking weights fully into account results in a generalized global assortativity coefficients of $r^\omega_{(1,1)} = 0.1928$, such that scientists tend to collaborate with others who are similar in terms of the number of papers they have been co-authors of (strength).

The first panel of \Cref{tab_generalised_local_assortativity_analysis_real_networks} presents the results of the global and local assortativity analysis of this network. Obviously, more than $70$ percent of the edges are assortative, since $P\big(\rho^\omega_e > 0\big) > 0.7$, for all parameter combinations $(\alpha, \beta)$. The proportion of assortative vertices is of a similar order of magnitude, with the proportion of $P\big(\rho^\omega_v > 0\big) = 68.9$ percent for the combination $(1,0)$. 

\cite{U.Pigorsch.2022} report for the NetScience network a consistently disassortative \textit{connection effect} and an inconsistent \textit{amplification effect}, since it is disassortative if degrees are used as vertex values but assortative if strengths are used.\footnote{They refer to a consistent connection effect or amplification effect if the resulting change of a network's value of the generalized assortativity coefficient is in the same direction when changing the parameters $\alpha$ and $\beta$, respectively. If changing the parameters results in changes of values of the generalized assortativity coefficient that work in opposite directions the effects are referred to as inconsistent.} Interestingly, both effects can also be assessed by comparing the average absolute magnitude of assortative and disassortative edges. This ratio changes if we use weighted rather than unweighted vertex values, i.e., excess strengths instead of excess degrees. For example, for the parameter combinations $(0,0)$ and $(0,1)$ the average absolute magnitude of assortative edges is higher than that for disassortative edges. On the other hand, focusing on the combinations $(1,0)$ and $(1,1)$, we find that the average absolute magnitude of disassortative edges is greater than that of assortative edges. This appears to be an indication of the connection effect. In contrast, if we compare the average absolute magnitudes of assortative and disassortative edges for a given value of the parameter $\alpha$, the amplification effect becomes apparent. For example, for $\alpha = 0$, the difference in the average absolute magnitudes between assortative and disassortative edges is reduced if $\beta = 1$ instead of $\beta =0$, since the magnitude of assortative edges is reduced in this case. For $\alpha = 1$, on the other hand, the difference in the average absolute magnitudes is reduced if $\beta = 1$ instead of $\beta =0$, because the magnitude of assortative edges increases more than for disassortative edges. The above confirms the analysis of \cite{U.Pigorsch.2022}.

\begin{table*}[!htbp]
	\centering
	\footnotesize
	\begin{tabular*}{\textwidth}{l @{\extracolsep{\fill}} cccc}
		\toprule
		\multirow{2}[2]{*}{Measure} & \multicolumn{2}{c}{$\alpha = 0$} & \multicolumn{2}{c}{$\alpha = 1$}\\ \cmidrule(lr){2-3}\cmidrule(lr){4-5}
		&
		\multicolumn{1}{c}{$\beta = 0$} & \multicolumn{1}{c}{$\beta = 1$} &
		\multicolumn{1}{c}{$\beta = 0$} & \multicolumn{1}{c}{$\beta = 1$} \\
		\midrule
		\addlinespace
		\multicolumn{5}{c}{NetScience} \\
		\addlinespace
		\midrule
		$r^\omega$& 0.462& 0.340& 0.102& 0.193 \\
	  	$P\big(\rho^\omega_e > 0\big)$& 0.772& 0.772& 0.713& 0.749 \\
	  	$\overline{\big(\rho^\omega_e\big)_+}$& 2.76e-04& 2.18e-04& 1.16e-04& 1.55e-04 \\
	  	$\overline{\big(\rho^\omega_e\big)_-}$& 1.96e-04& 1.93e-04& 1.59e-04& 1.83e-04 \\
	  	$P\big(\rho^\omega_v > 0\big)$& 0.755& 0.737& 0.689& 0.738 \\
		\midrule
		\addlinespace
		\multicolumn{5}{c}{C. Elegans} \\
		\addlinespace
		\midrule
		\multicolumn{5}{l}{out--in} \\
		\midrule
		$r^\omega$&-0.233&-0.355&-0.181&-0.292 \\
		$P\big(\rho^\omega_e > 0\big)$& 0.511& 0.494& 0.610& 0.633 \\
		$\overline{\big(\rho^\omega_e\big)_+}$& 1.19e-04& 1.18e-04& 6.18e-05& 8.21e-05 \\
		$\overline{\big(\rho^\omega_e\big)_-}$& 3.27e-04& 4.14e-04& 2.94e-04& 4.80e-04 \\
		$P\big(\rho^\omega_v > 0\big)$& 0.306& 0.350& 0.455& 0.495 \\
		\midrule
		\multicolumn{5}{l}{out--out} \\
		\midrule
		$r^\omega$& 0.099& 0.269& 0.065& 0.148 \\
		$P\big(\rho^\omega_e > 0\big)$& 0.562& 0.564& 0.591& 0.593 \\
		$\overline{\big(\rho^\omega_e\big)_+}$& 2.74e-04& 3.50e-04& 1.99e-04& 2.24e-04 \\
		$\overline{\big(\rho^\omega_e\big)_-}$& 2.55e-04& 1.89e-04& 2.20e-04& 1.71e-04 \\
		$P\big(\rho^\omega_v > 0\big)$& 0.643& 0.640& 0.646& 0.640 \\
		\midrule
		\multicolumn{5}{l}{in-in} \\
		\midrule
		$r^\omega$&-0.092&-0.132&-0.068&-0.098 \\
		$P\big(\rho^\omega_e > 0\big)$& 0.572& 0.677& 0.694& 0.734 \\
		$\overline{\big(\rho^\omega_e\big)_+}$& 1.14e-04& 1.12e-04& 6.14e-05& 9.66e-05 \\
		$\overline{\big(\rho^\omega_e\big)_-}$& 2.44e-04& 4.08e-04& 2.34e-04& 4.23e-04 \\
		$P\big(\rho^\omega_v > 0\big)$& 0.549& 0.616& 0.582& 0.636 \\
		\midrule
		\multicolumn{5}{l}{in--out} \\
		\midrule
		$r^\omega$&-0.026& 0.138& 0.061& 0.125 \\
		$P\big(\rho^\omega_e > 0\big)$& 0.531& 0.530& 0.645& 0.658 \\
		$\overline{\big(\rho^\omega_e\big)_+}$& 2.28e-04& 3.19e-04& 1.66e-04& 1.92e-04 \\
		$\overline{\big(\rho^\omega_e\big)_-}$& 2.82e-04& 2.35e-04& 2.27e-04& 2.13e-04 \\
		$P\big(\rho^\omega_v > 0\big)$& 0.532& 0.549& 0.660& 0.653 \\
	\bottomrule
	\end{tabular*}
	\caption[Generalized local assortativity analysis of weighted real-world networks for different parameter combinations $(\alpha,\beta)$.]{\textbf{Generalized local assortativity analysis of real-world networks for different parameter combinations $(\alpha,\beta)$.} Reported are the generalized assortativity coefficient $r^\omega$, fraction of local assortative edges $P(\rho^\omega_e > 0)$, average absolute magnitude of assortative edges $\overline{(\rho^\omega_e)_+}$, average absolute magnitude of disassortative edges $\overline{(\rho^\omega_e)_-}$, fraction of local assortative vertices $P(\rho^\omega_v > 0)$. Considered are the undirected \textit{NetScience} network (a co-authorship network of scientists working on network theory \citep{Newman.2001b}, in which authors are connected if they have co-authored one or more papers); and the directed neural network of the nematode worm \textit{C. Elegans} (where edges indicate that two neurons are connected by either a synapse or a gap junction \citep{Watts.1998, White.1986}).
	}
	\label{tab_generalised_local_assortativity_analysis_real_networks}
\end{table*}

\begin{figure*}[!htb]
	\centering
	\begin{subfigure}{.48 \textwidth}
		\centering
		\caption{\label{fig_netscience_a}}
		\includegraphics[width = \textwidth]{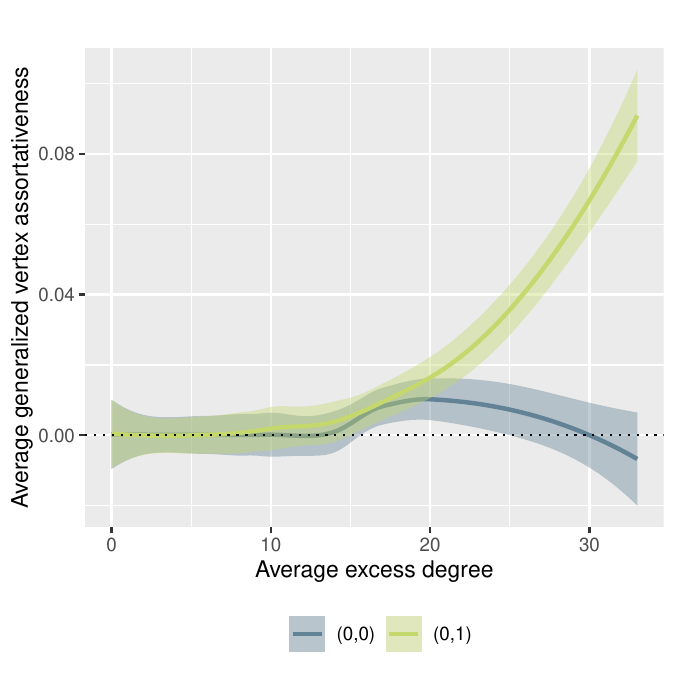}
	\end{subfigure}
	\hfill
	\begin{subfigure}{.48 \textwidth}
		\centering
		\caption{\label{fig_netscience_b}}
		\includegraphics[width = \textwidth]{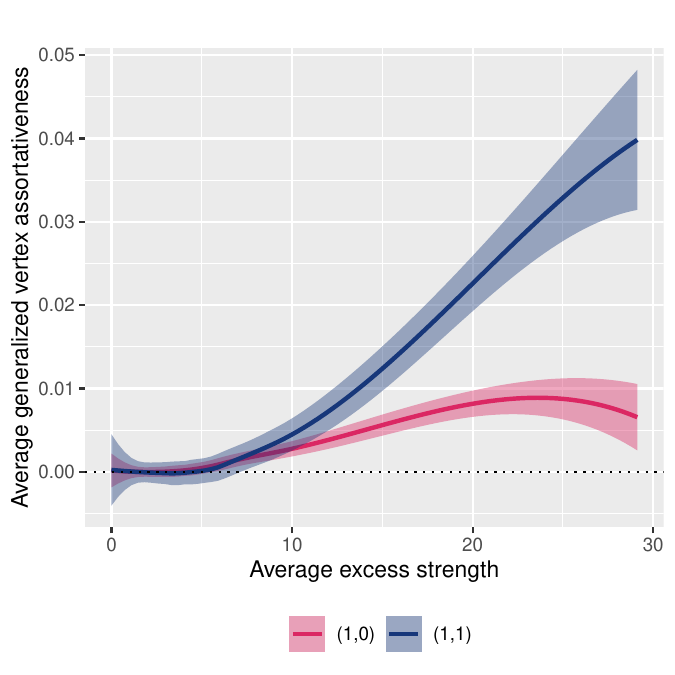}
	\end{subfigure}
	\hfill
	\begin{subfigure}{.48 \textwidth}
		\centering
		\caption{\label{fig_netscience_c}}
		\includegraphics[width = \textwidth]{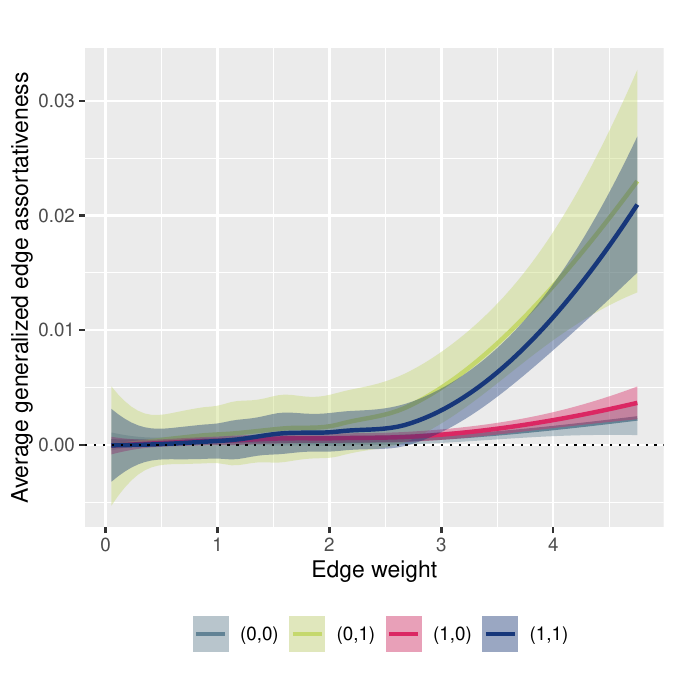}
	\end{subfigure}
	\caption[Generalized vertex and edge assortativeness profiles of the \textit{NetScience} network.]{\textbf{Generalized vertex and edge assortativeness profiles of the \textit{NetScience} network.} Generalized vertex degree assortativeness (a), strength assortativeness (b) and edge assortativeness (c) profiles of the \textit{NetScience} network for all four parameter combinations $(\alpha, \beta)$. The profiles are obtained by smoothing the data with loess regression (the shaded area indicates 95 percent confidence bands).}
	\label{plt_netscience_local_vertex_assortativity}
\end{figure*}
\Cref{fig_netscience_a,fig_netscience_b}, depict the vertex assortativeness values plotted against the average excess degree or strength of a vertex, respectively. Moreover, \Cref{fig_netscience_c} shows the edge assortativeness values plotted against the corresponding weight of an edge. Analyzing \Cref{fig_netscience_a,fig_netscience_b}, for the combinations $(0,1)$ and $(1,1)$, reveals that the higher the average excess degree or strength of a vertex, the more assortative it is. For combinations $(0,0)$ and $(1,0)$, vertices tend to be non-assortative up to a medium average excess degree or strength, then assortative, and then non-assortative again. In terms of content, there seems to be a certain degree or strength where authors prefer to form connections with other equally well-connected authors. Below or above, research collaborations seem to be of a more disassortative nature, both in terms of degrees and in terms of strength. However, recurring collaborations tend to exist between well-connected authors, as is indicated by the slope of the curves when the weighted correlation is computed, i.e. for $\beta=1$.

\Cref{fig_netscience_c} shows that edges tend to be more assortative the greater the edge weight, with this effect being stronger in the case of degrees than in the case of strengths. Again, the inconsistent amplification effect can be observed since the curve for the combination $(0,1)$ is above that of $(0,0)$, whereas for $(1,1)$ it is below that for $(1,0)$. 
Oftentimes it may also be of interest to examine which vertices or edges are particularly assortative or disassortative. For the network under consideration a particular assortative vertex corresponds to an author collaborating usually with others that have a similar number of co-authors (degree assortative) or a similar number of co-authored publications (strength assortative). Disassortative vertices, on the other hand, correspond to authors that usually collaborate with others who are unlike them, i.e., authors with many co-authors (degree disassortative) or co-authored publications (strength disassortative) tend to collaborate with others that have few co-authors or few co-authored publications, respectively, and vice versa. Assortative edges connect authors with a similar number of co-authors (degree assortative) or co-authored publications (strength assortative), whereas disassortative edges connect authors that are dissimilar. \Cref{tab_local_vertex_assortativity_rank_analysis,tab_local_edge_assortativity_rank_analysis} in the Appendix list the top assortative and disassortative vertices and edges.

\subsubsection{Caenorhabditis Elegans Neural Network}

The neural network of the nematode worm \textit{Caenorhabditis elegans} (C. Elegans) is an example of a completely mapped neural network, cf. \cite{Watts.1998} and \cite{White.1986}. A node in the directed and weighted network represents a neuron and an edge between two neurons indicates that they are connected by either a synapse or a gap junction. However, we could not find any information on how edge weights are defined in this network, so we must assume that they somehow reflect the cost or capacity of communication between the neurons (e.g., distance, speed, volume, or bandwidth) as it is the usual way to define edge weights in brain networks of that type, cf. \cite{Faskowitz.2022}.

The global assortativity coefficient $r_{(\alpha,\beta)}^{\omega}$ reported in \Cref{tab_generalised_local_assortativity_analysis_real_networks} indicates that the \textit{C. Elegans} network is globally disassortative for the modes \textit{out--in} and \textit{in--in} for all combinations $(\alpha, \beta)$. Moreover, the network is globally assortative for the modes \textit{out--out} and \textit{in--out} for all combinations $(\alpha, \beta)$, except for the mode \textit{in--out} for the combination $(0,0)$, for which the network is also disassortative. \Cref{tab_generalised_local_assortativity_analysis_real_networks} also presents the results of the local assortativity measures of the \textit{C. Elegans} network. The reported measures allow to conduct for each mode a local analysis along the lines of our discussion of the local assortativity structure of the \textit{NetScience} network. However, for the sake of brevity, we refrain from doing so and, instead, focus directly on the analysis of the local vertex and edge assortativeness profiles.

\begin{figure*}[!htb]
	\centering
	\includegraphics[width = .85\linewidth]{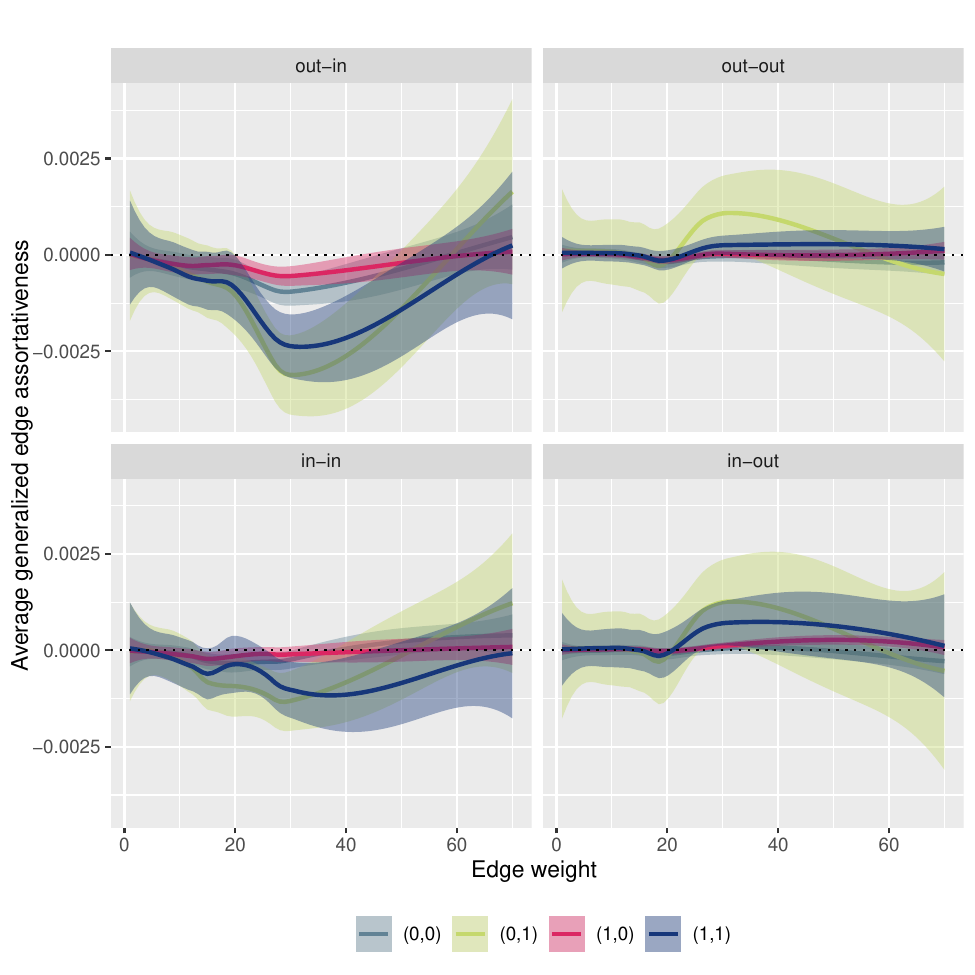}
	\caption[Generalized edge assortativeness profiles of the \textit{C. Elegans} network for all four modes of assortativity.]{\textbf{Generalized edge assortativeness profiles of the \textit{C. Elegans} network for all four modes of assortativity.} Generalized edge assortativeness profiles of the \textit{C. Elegans} neural network for all four parameter combinations $(\alpha, \beta)$. The profiles are obtained by smoothing the data with loess regression (the shaded area indicates 90 percent confidence bands).}
	\label{plt_celegansneural_local_edge_assortativity}
\end{figure*}
\begin{figure*}[!htb]
	\centering
	\includegraphics[width = .85\linewidth]{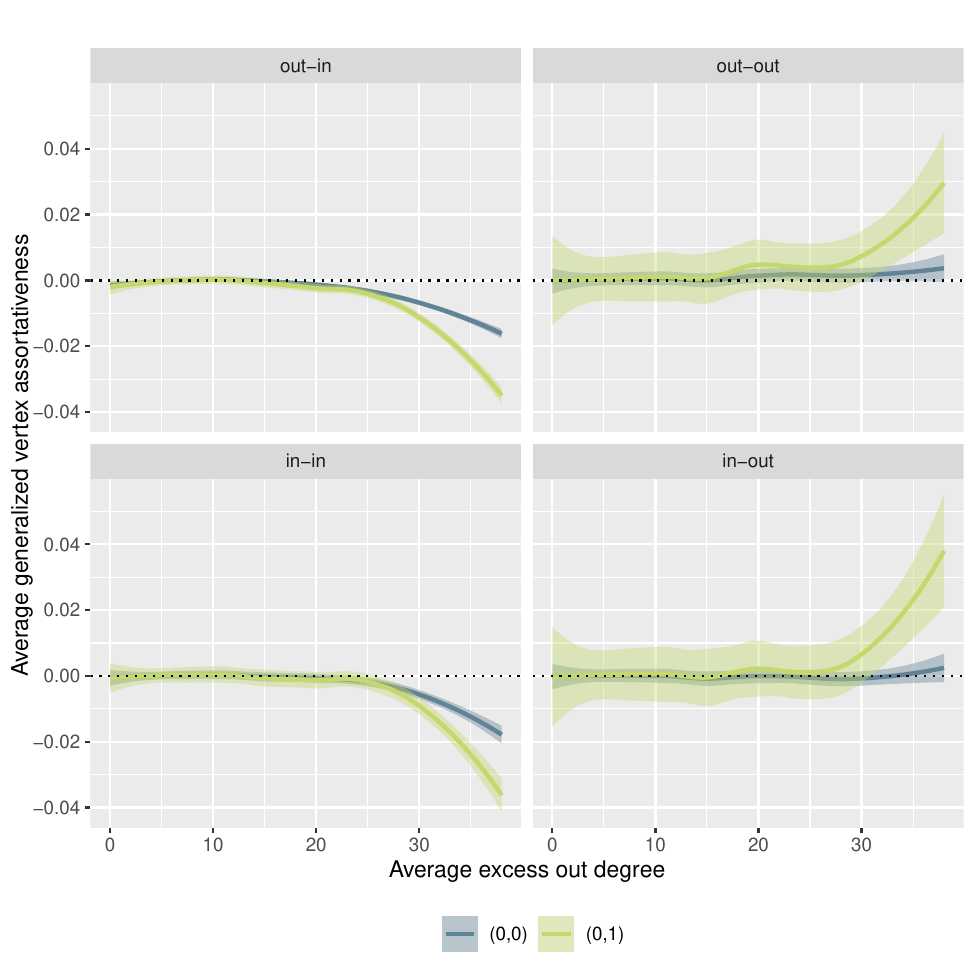}
	\caption[Generalized vertex degree assortativeness profiles of the \textit{C. Elegans} network for all four modes of assortativity.]{\textbf{Generalized vertex degree assortativeness profiles of the \textit{C. Elegans} network for all four modes of assortativity.} Generalized vertex degree assortativeness profiles of the \textit{C. Elegans} neural network using excess degrees as vertex values ($\alpha=0$). The profiles are obtained by smoothing the data with loess regression (the shaded area indicates 95 percent confidence bands).}
	\label{plt_celegansneural_local_vertex_assortativity_out_degree}
\end{figure*}
\begin{figure*}[!htb]
	\centering
4	\includegraphics[width = .85\linewidth]{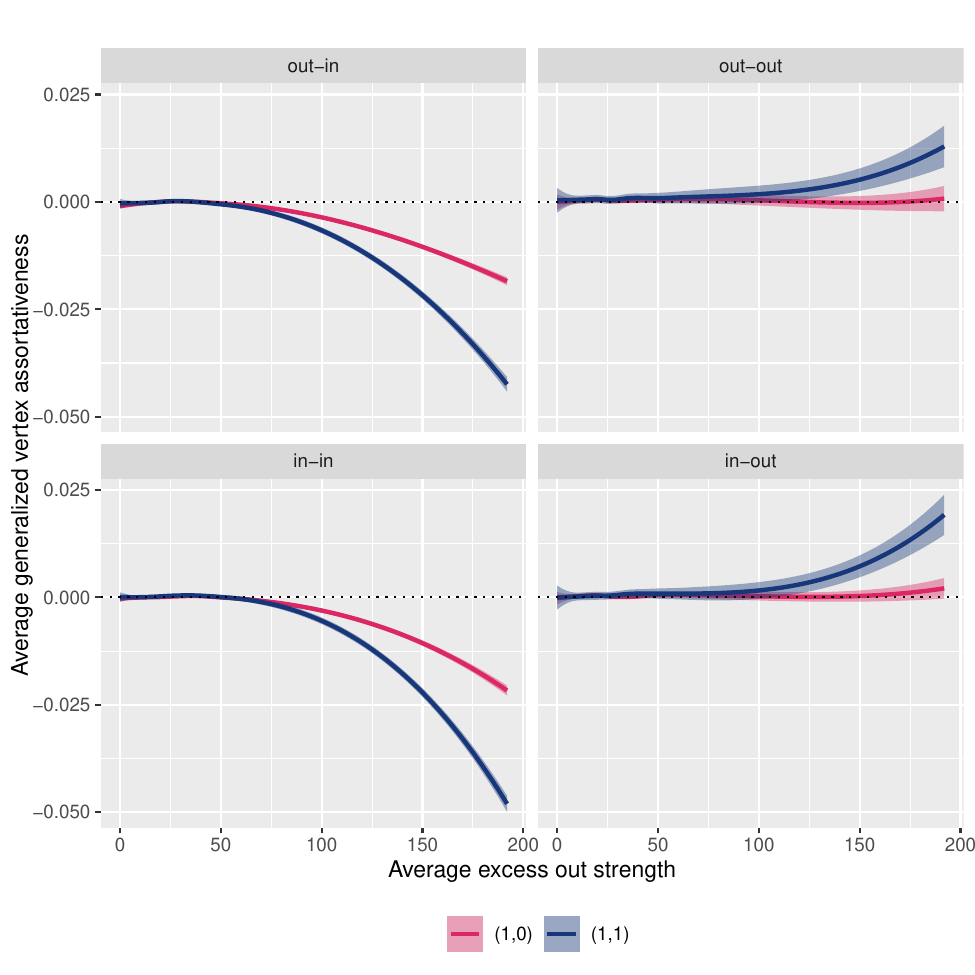}
	\caption[Generalized vertex strength assortativeness profiles of the \textit{C. Elegans} network for all four modes of assortativity.]{\textbf{Generalized vertex strength assortativeness profiles of the \textit{C. Elegans} network for all four modes of assortativity.} Generalized vertex strength assortativeness profiles of the \textit{C. Elegans} neural network using excess strengths as vertex values ($\alpha=1$). The profiles are obtained by smoothing the data with loess regression (the shaded area indicates 95 percent confidence bands).}
	\label{plt_celegansneural_local_vertex_assortativity_out_strength}
\end{figure*}

\Cref{plt_celegansneural_local_edge_assortativity} shows the average edge assortativeness values plotted against the corresponding edge weight. In \Cref{plt_celegansneural_local_vertex_assortativity_out_degree,plt_celegansneural_local_vertex_assortativity_out_strength} the average vertex assortativeness values are plotted against the corresponding average excess out-degree or -strength of a vertex. For reasons of clarity, we consider the averages of the edge and vertex assortativeness values for this network. 

Considering the edge assortativeness values first, we see that the modes \textit{out--in} and \textit{in--in} as well as \textit{out--out} and \textit{in--out} turn out be structurally quite similar, respectively. For the \textit{out--in} mode, edges with low and high weights tend to be non-assortative. For medium-sized weights edges tend to be disassortative and the \textit{out--in} edge assortativeness values decrease more clearly for the combinations $(0,1)$ and $(1,1)$ than for the combinations $(0,0)$ and $(1,0)$. This is again an indication of the amplification effect, which is consistent since the curve for $(0,1)$ is below that of $(0,0)$ and that for $(1,1)$ is below that of $(1,0)$. For the \textit{in--in} mode we find a similar, but less pronounced local edge assortativeness structure, whereas for the \textit{out--out} mode, the edge assortativeness values tend to be non-assortative for all combinations of $(\alpha, \beta)$, except for $(0,1)$. Over the middle edge weight range the edges tend to be assortative. In addition, for the structurally similar \textit{in--out} mode, the curve bends slightly in the area of middle edge weights especially for the parameter combination $(1,1)$.

By considering \Cref{plt_celegansneural_local_vertex_assortativity_out_degree,plt_celegansneural_local_vertex_assortativity_out_strength}, we see that with increasing average excess out-degree or -strength, vertices tend to become more disassortative for the \textit{out--in} and \textit{in--in} modes of assortativity. Conversely, for the modes \textit{out--out} and \textit{in--out}, vertices tend to be more assortative with increasing average excess out-degree or -strength. However, for all modes, the pattern is more pronounced when the weighted correlation is considered, i.e., $\beta = 1$ instead of $\beta = 0$.

The local assortativity analysis revealed how the global assortativity structure of a network is composed. The \textit{C. Elegans} network is \textit{out--in} disassortative because edges with a mid-range edge weight are disassortative, while low-weighted edges and high-weighted edges tend to be non-assortative (see \Cref{plt_celegansneural_local_edge_assortativity}). The \textit{NetScience} network, on the other hand, is assortative, since edges tend to be more assortative the higher the edge weight. The effect is more or less pronounced for both networks, depending on whether degrees or strengths are used as vertex values. 

Such a detailed analysis has only become feasible through the introduction of our generalized local assortativity measures, which help to further decompose the assortativity structure of a network. For a comparison across networks, this also allows to further differentiate the topology of networks that exhibit a similar global assortativity, as we have shown in the analysis of the WRG and WSF models. In summary, the generalized local assortativity analysis provides additional valuable information about the assortativity of a network. Importantly, such information can now be computed also for weighted networks.

\section{Discussion and Future Work}\label{sec_discussion_local}

In this paper we generalized local assortativity to weighted networks. In particular, we showed the equivalence of two popular local assortativity approaches and derived several measures that allow to assess the assortativeness of individual edges, vertices or components not only of unweighted, but importantly, also of weighted networks. We demonstrated their usefulness by analyzing the local assortativity structure of theoretical and real-world networks. Along the way, we also proposed local assortativity profiles, which are informative about the pattern of local assortativity either with respect to edge weight or vertex strength. Such profiles have been analyzed in \cite{Piraveenan.2008} for the unweighted vertex-based local assortativity measure. Since our generalized local assortativity can be either vertex-based or edge-based, we extended the assortativity profiles accordingly providing additional information when considering weighted networks.

\cite{Thedchanamoorthy.2014} suggest an allegedly computationally less expensive alternative definition of local assortativity that, contrary to the definition by \cite{Piraveenan.2008, Piraveenan.2010}, does not pivot on the global mean excess degree (or strength) of the ends of an edge, $\bar{U}_{q_k}$. They argue, that their approach is computationally less expensive and that the definition of \cite{Piraveenan.2008, Piraveenan.2010} is counter-intuitive, as an edge, which connects two vertices with different degrees, but both higher than $\bar{U}_{q_k}$, is considered assortative, whereas an edge that connects two vertices with similar degrees, where one has a degree higher than $\bar{U}_{q_k}$, and the other one has a degree lower than $\bar{U}_{q_k}$, is considered disassortative. However, unlike that of \cite{Thedchanamoorthy.2014} the local assortativity measure of \cite{Piraveenan.2008, Piraveenan.2010} is in line with the original and widely accepted definition of assortativity going back to \cite{Newman.2002}. Moreover, it is not computationally more expensive, as it can be directly computed once the global assortativity coefficient is calculated. In this paper, we therefore did not consider the approach by \cite{Thedchanamoorthy.2014}, but instead adopted the framework of \cite{Piraveenan.2008, Piraveenan.2010} resulting in generalized local assortativity measures that are consistent with the common definition of assortativity and computationally less expensive.  
Moreover, as a consequence, our generalized edge assortativeness values, $\rho_e^\omega(\alpha, \beta, \textit{mode})$, are directly available when computing the assortativity coefficient of a network, and thus, our definition of local assortativity does not bear any additional computational costs.

Note that analyzing the local assortativity structure of a network via the generalized edge assortativeness values, $\rho^\omega_e(\alpha, \beta, \textit{mode})$, captures only the direct contribution of an edge $e$ to the global assortativity coefficient.
More precisely, as the generalized assortativeness values are defined as a product of the scaled differences of the excess degrees (or strengths) of both ends of an edge and their respective means, edges are considered in isolation.
However, if a vertex $u$ is incident with more than one edge, then, the presence of edge $e$ increases the excess degree (or strength) of that vertex when considering any edge other than $e$, which can be seen as an indirect contribution to the global assortativity.
This potential indirect contribution is neglected in the definition of the generalized assortativeness values, but is also neglected in the previous local assortativity definitions by \cite{Piraveenan.2008,Piraveenan.2010,Piraveenan.2012} and \cite{Thedchanamoorthy.2014}.
In order to decide if this is a drawback, a more in-depth analysis is necessary. We plan on revisiting this topic in future research where we will compare the generalized assortativeness values to an alternative local edge assortativity measure that also captures indirect contributions to the global assortativity. For example, an unsophisticated alternative local edge assortativity measure is the following:
\begin{align}
	\rho_e^J = \tilde{r}\frac{\Delta\tilde{r}(e)}{\sum_{i=1}^M \Delta\tilde{r}(i)},
	\label{eq_jackknife_local_edge_assortativeness}
\end{align}
where $\tilde{r}$ is any global assortativity coefficient, such as e.g. the generalized assortativity coefficient, $\tilde{r} = r^\omega_{(\alpha, \beta)}$. Then $\Delta\tilde{r}(e) = \tilde{r} - \tilde{r}_{(-e)}$ is the difference between the global assortativity $\tilde{r}$ of a network and the jackknife statistic $\tilde{r}_{(-e)}$, i.e., the global assortativity of the same network but with the $e$-th edge removed.
Apparently, the measure $\rho_e^J$ is based on the jackknife method, which is indicated by its superskript $J$. We refer to it as the jackknife local (edge) assortativeness values. It captures the direct as well as the above-mentioned indirect contribution of an edge $e$ to the global assortativity of a network. However, its computational costs are considerably larger as compared to the generalized local assortativeness  (see \cref{eq_rho_directed_weighted_alpha_beta}). In fact, it increases by a factor of $(M + 1)$. Nevertheless, for small- and medium-sized networks, the jackknife local edge assortativeness values can serve as a benchmark for determining the accuracy of the generalized local assortativeness values. So far, we conjecture that the indirect influence on the assortativity of an edge becomes negligible the bigger the network is, which leads us to believe that the use of the generalized local assortativeness values for measuring the local edge assortativity is appropriate in most cases, especially when analyzing very large real world networks, for which computing the local assortativity based on the jackknife statistics would result in large computational costs.

Moreover, we expect that the generalized local assortativity can be extended to the more complex definitions of assortativity of \cite{Meghanathan.2016} or \cite{Arcagni.2017,Arcagni.2021}, as they are, at their core, also based on Pearson's correlation.

For future research it will also be interesting to assess the significance of local assortativity profiles. Since local assortativity is a \textit{third-order} graph metric this would require a feature preserving graph rewiring algorithm in order to decide whether the observed local assortativity profile significantly deviates from one of a suitable null model (either generative or by link-rewiring). However, such a rewiring algorithm would have to preserve also the global assortativity of the network, and thus, the link-rewiring algorithm of \cite{Rubinov.2011}, which was employed in \cite{U.Pigorsch.2022} in order to assess the significance of the global generalized assortativity coefficient, is not applicable here as it only preserves the degree and strength distribution of the observed network. Unfortunately, to the best of our knowledge, such a null model of weighted networks that preserves the observed generalized assortativity does not yet exist. 

In future research it may also be interesting to apply our local assortativity measures to study epidemic spreading. \cite{Wu.2005} analyze the properties of weighted scale free networks, in particular, the epidemic spreading process using an \textit{susceptible-infected} (SI) model. Alternatively, epidemic spreading could be analyzed in scale free networks that also show assortative or disassortative mixing, e.g. by considering the \textit{mutual attraction} model introduced by \cite{Wang.2006} for generating assortative and disassortative networks. It will be interesting to assess whether a network's resilience against exogenous shocks can be increased by removing edges with certain local assortativeness values, such that the epidemic spreading process could be slowed down. Based on such findings, policy advice can be given, which can also be interesting for financial networks such as the cryptocurrency network. 

\section*{Acknowledgments}

The authors thank the participants of the $ 11^{\text{th}} $ International Conference on Complex Networks and Their Applications (Complex Networks 2022), and the participants of the $ 15^{\text{th}} $
International Conference of the ERCIM WG on Computational and Methodological Statistics (CMStatistics 2022) for helpful discussions and comments.

\section*{Funding Statement}
This research did not receive any specific grant from funding agencies in the public, commercial, or not-for-profit sectors.

\printbibliography 


\appendix
\section{Appendix}
\label{sec:sample:appendix}

\begin{table*}[!htb]
	\centering
	\scriptsize
	\begin{tabular*}{\textwidth}{r @{\extracolsep{\fill}} lcclcc}
		\toprule
		& \multicolumn{3}{c}{$(\alpha = 0, \beta = 0)$} & \multicolumn{3}{c}{$(\alpha = 1, \beta = 1)$} \\ \cmidrule(lr){2-4}\cmidrule(lr){5-7}
	   \multicolumn{1}{r}{Rank} & \multicolumn{1}{l}{Name} & \multicolumn{1}{c}{$\rho^\omega_v$} & \multicolumn{1}{r}{Degree} & \multicolumn{1}{l}{Name} & \multicolumn{1}{c}{$\rho^\omega_v$} & \multicolumn{1}{r}{Strength} \\
	   \midrule

	   1&UETZ, P& 0.0159&20&JEONG, H& 0.0252&18 \\
	   2&CAGNEY, G& 0.0159&20&BARABASI, A& 0.0226&30 \\
	   3&MANSFIELD, T& 0.0159&20&PASTORSATORRAS, R& 0.0171&17 \\
	   4&GIOT, L& 0.0150&19&VESPIGNANI, A& 0.0119&15 \\
	   5&JUDSON, R& 0.0150&19&OLTVAI, Z& 0.0069&10 \\
	   \addlinespace
	   6&KNIGHT, J& 0.0150&19&MORENO, Y& 0.0062&15 \\
	   7&LOCKSHON, D& 0.0150&19&VAZQUEZ, A& 0.0062&11 \\
	   8&NARAYAN, V& 0.0150&19&VICSEK, T& 0.0046& 9 \\
	   9&SRINIVASAN, M& 0.0150&19&ALBERT, R& 0.0037& 8 \\
	   10&POCHART, P& 0.0150&19&SOLE, R& 0.0035&15 \\
	   \addlinespace
	   11&QURESHIEMILI, A& 0.0150&19&WATTS, D& 0.0029& 9 \\
	   12&LI, Y& 0.0150&19&BARTHELEMY, M& 0.0021& 9 \\
	   13&GODWIN, B& 0.0150&19&HILGETAG, C& 0.0021&11 \\
	   14&CONOVER, D& 0.0150&19&DIAZGUILERA, A& 0.0020&11 \\
	   15&KALBFLEISCH, T& 0.0150&19&NEWMAN, M& 0.0019&23 \\
	   \addlinespace
	   16&VIJAYADAMODAR, G& 0.0150&19&GUIMERA, R& 0.0019&10 \\
	   17&YANG, M& 0.0150&19&KAHNG, B& 0.0018&11 \\
	   18&JOHNSTON, M& 0.0150&19&STROGATZ, S& 0.0016& 8 \\
	   19&FIELDS, S& 0.0150&19&YOUNG, M& 0.0015&13 \\
	   20&ROTHBERG, J& 0.0150&19&HOLME, P& 0.0014& 9 \\
	   $\vdots$ &&& &&& \\
	   1442&LEICHT, E&-0.0007& 2&ALON, U&-0.0006& 8 \\
	   1443&MACDONALD, P&-0.0008& 2&MACDONALD, P&-0.0007& 1 \\
	   1444&BIANCONI, G&-0.0008& 4&SMITH, E&-0.0007& 2 \\
	   1445&DOBRIN, R&-0.0008& 3&FERRERICANCHO, R&-0.0007& 4 \\
	   1446&BEG, Q&-0.0008& 3&VALVERDE, S&-0.0007& 5 \\
	   \addlinespace
	   1447&HU, G&-0.0008&11&PACHECO, A&-0.0007& 4 \\
	   1448&KOVACS, B&-0.0008& 4&MASON, S&-0.0007& 1 \\
	   1449&LATORA, V&-0.0008&15&WUCHTY, S&-0.0008& 1 \\
	   1450&GLOT, L&-0.0008& 3&HUBERMAN, B&-0.0008& 8 \\
	   1451&SOLE, R&-0.0009&17&BOCCALETTI, S&-0.0009&12 \\
	   \addlinespace
	   1452&DEZSO, Z&-0.0009& 1&BIANCONI, G&-0.0009& 3 \\
	   1453&YOOK, S&-0.0009& 4&LATORA, V&-0.0009&11 \\
	   1454&TU, Y&-0.0009& 3&BORNHOLDT, S&-0.0010& 8 \\
	   1455&BOCCALETTI, S&-0.0011&19&YOOK, S&-0.0011& 3 \\
	   1456&WUCHTY, S&-0.0011& 2&GASTNER, M&-0.0011& 1 \\
	   \addlinespace
	   1457&DIAZGUILERA, A&-0.0012&15&LUSSEAU, D&-0.0011& 1 \\
	   1458&MASON, S&-0.0013& 3&PARK, J&-0.0011& 1 \\
	   1459&YOUNG, M&-0.0016&20&MONTOYA, J&-0.0011& 2 \\
	   1460&BARABASI, A&-0.0022&34&DEZSO, Z&-0.0015& 1 \\
	   1461&NEWMAN, M&-0.0059&27&GIRVAN, M&-0.0020& 3 \\

	   \bottomrule
	\end{tabular*}
	\caption[Generalized vertex assortativeness ranking (\textit{NetScience} network).]{\textbf{Generalized vertex assortativeness ranking of the \textit{NetScience} network.} Reported are the 20 most assortative as well as the 20 most disassortative actors of the network, with respect to generalized vertex assortativeness, $\rho^\omega_v(\alpha, \beta)$, for the two parameter combinations $(\alpha, \beta) = \{(0,0), (1,1)\}$.}
	\label{tab_local_vertex_assortativity_rank_analysis}
\end{table*}

\begin{sidewaystable*}[!htb]
	\centering
	\tiny
	\begin{tabular*}{\textwidth}{r @{\extracolsep{\fill}} llcccccllccccc}
		\toprule
		& \multicolumn{7}{c}{$(\alpha = 0, \beta = 0)$} & \multicolumn{7}{c}{$(\alpha = 1, \beta = 1)$} \\ \cmidrule{2-8}\cmidrule{9-15}
	   \multicolumn{1}{r}{Rank} & \multicolumn{1}{l}{$u$} & \multicolumn{1}{l}{$v$} & \multicolumn{1}{c}{$\rho^\omega_{e_{uv}}$} & \multicolumn{1}{r}{$k^\prime_u$} & \multicolumn{1}{r}{$k^\prime_v$} & \multicolumn{1}{r}{$\bar{k}_u$} & \multicolumn{1}{r}{$\bar{k}_v$} & \multicolumn{1}{l}{$u$} & \multicolumn{1}{l}{$v$} & \multicolumn{1}{c}{$\rho^\omega_{e_{uv}}$} & \multicolumn{1}{r}{$s^\prime_u$} & \multicolumn{1}{r}{$s^\prime_v$} & \multicolumn{1}{r}{$\bar{s}_u$} & \multicolumn{1}{r}{$\bar{s}_v$} \\
	   \midrule

	   1&BARABASI, A&JEONG, H& 0.0058&34&27&33&26&BARABASI, A&JEONG, H& 0.0385&30&18&29.1176&17.3333 \\
	   2&BARABASI, A&OLTVAI, Z& 0.0041&34&21&33&20&PASTORSATORRAS, R&VESPIGNANI, A& 0.0118&17&15&15.8667&13.9286 \\
	   3&JEONG, H&OLTVAI, Z& 0.0030&27&21&26&20&BARABASI, A&OLTVAI, Z& 0.0107&30&10&29.1176& 9.5238 \\
	   4&BARABASI, A&VICSEK, T& 0.0026&34&16&33&15&PASTORSATORRAS, R&SOLE, R& 0.0084&17&15&15.8667&14.1176 \\
	   5&NEWMAN, M&SOLE, R& 0.0022&27&17&26&16&NEWMAN, M&SOLE, R& 0.0078&23&15&22.1481&14.1176 \\
	   \addlinespace
	   6&JEONG, H&VICSEK, T& 0.0019&27&16&26&15&BARABASI, A&VICSEK, T& 0.0072&30& 9&29.1176& 8.4375 \\
	   7&UETZ, P&CAGNEY, G& 0.0018&20&20&19&19&VESPIGNANI, A&MORENO, Y& 0.0064&15&15&13.9286&13.9286 \\
	   8&UETZ, P&MANSFIELD, T& 0.0018&20&20&19&19&PASTORSATORRAS, R&MORENO, Y& 0.0057&17&15&15.8667&13.9286 \\
	   9&CAGNEY, G&MANSFIELD, T& 0.0018&20&20&19&19&NEWMAN, M&WATTS, D& 0.0055&23& 9&22.1481& 7.7143 \\
	   10&GIOT, L&UETZ, P& 0.0017&19&20&18&19&MORENO, Y&VAZQUEZ, A& 0.0050&15&11&13.9286&10.0833 \\
	   $\vdots$ &&&&&&&&&&&&&& \\
	   2733&BARABASI, A&DOBRIN, R&-0.0011&34& 3&33& 2&MORENO, Y&PACHECO, A&-0.0013&15& 4&13.9286& 3.0000 \\
	   2734&BARABASI, A&BEG, Q&-0.0011&34& 3&33& 2&SOLE, R&FERRERICANCHO, R&-0.0013&15& 4&14.1176& 3.0000 \\
	   2735&BARABASI, A&MASON, S&-0.0011&34& 3&33& 2&BARABASI, A&BIANCONI, G&-0.0017&30& 3&29.1176& 2.2500 \\
	   2736&BARABASI, A&TU, Y&-0.0011&34& 3&33& 2&BARABASI, A&YOOK, S&-0.0017&30& 3&29.1176& 2.2500 \\
	   2737&NEWMAN, M&GASTNER, M&-0.0013&27& 1&26& 0&NEWMAN, M&GASTNER, M&-0.0021&23& 1&22.1481& 0.0000 \\
	   \addlinespace
	   2738&NEWMAN, M&LUSSEAU, D&-0.0013&27& 1&26& 0&NEWMAN, M&LUSSEAU, D&-0.0021&23& 1&22.1481& 0.0000 \\
	   2739&NEWMAN, M&PARK, J&-0.0013&27& 1&26& 0&NEWMAN, M&PARK, J&-0.0021&23& 1&22.1481& 0.0000 \\
	   2740&BARABASI, A&MACDONALD, P&-0.0014&34& 2&33& 1&SOLE, R&MONTOYA, J&-0.0022&15& 2&14.1176& 0.0000 \\
	   2741&BARABASI, A&WUCHTY, S&-0.0014&34& 2&33& 1&BARABASI, A&DEZSO, Z&-0.0029&30& 1&29.1176& 0.0000 \\
	   2742&BARABASI, A&DEZSO, Z&-0.0017&34& 1&33& 0&NEWMAN, M&GIRVAN, M&-0.0041&23& 3&22.1481& 1.5000 \\
	   \bottomrule
	\end{tabular*}
	\caption[Generalized edge assortativeness ranking (\textit{NetScience} network).]{\textbf{Generalized edge assortativeness ranking of the \textit{NetScience} network.} Reported are the 10 most assortative as well as the 10 most disassortative connections of the network, with respect to generalized edge assortativeness, $\rho^\omega_{e_{uv}}(\alpha, \beta)$ together with the mean average excess degree, $\bar{k}$, as well as the degree, $k\prime$, for the combination $(\alpha, \beta) = (0,0)$, and the mean average excess strength, $\bar{s}$, as well as the strength, $s\prime$, for the combination $(\alpha, \beta) = (1,1)$, for both ends, $u$ and $v$, of an edge $e$.}
	\label{tab_local_edge_assortativity_rank_analysis}
\end{sidewaystable*}

\end{document}